\pdfoutput=1

\documentclass[11pt,letterpaper]{article}
\renewcommand{\footnoterule}{%
  \kern -3pt
  \hrule width \textwidth height 0.5pt
  \kern 2pt
}

\usepackage{latexsym}
\usepackage{amssymb}
\usepackage{amsmath}
\usepackage{amsthm}
\usepackage{accents}
\usepackage{bbm}
\usepackage{setspace}
\usepackage{graphicx}
\usepackage{xcolor} 
\usepackage{bm}
\usepackage{enumerate}
\usepackage{sectsty}
\usepackage{subfig}
\usepackage{float}
\usepackage{caption}
\usepackage{wasysym}
\usepackage{times,fancyhdr}
\usepackage{mathrsfs}
\usepackage{hyperref}

\newtheorem*{definition}{Definition}

\newcommand{\PRLsep}{\noindent\makebox[\linewidth]{\resizebox{0.750\linewidth}{1pt}{$\blacklozenge$}}\bigskip}
\newcommand{\PRLsepsmall}{\noindent\makebox[\linewidth]{\resizebox{0.40\linewidth}{0.5pt}{$\blacklozenge$}}\bigskip}
\newcommand*{\Scale}[2][4]{\scalebox{#1}{$#2$}}%

\newcommand{\tinyrmsub}[1]{\mbox{{\tiny{#1}}}}
\newcommand{\rout}{r_{\tinyrmsub{out}}}
\newcommand{\rin}{r_{\tinyrmsub{in}}}
\newcommand{\gsm}{\tilde{\mu}(r)}
\newcommand{\gsl}{\tilde{\lambda}(r)}
\newcommand{\gsmrout}{\tilde{\mu}(\rout)}
\newcommand{\gslrout}{\tilde{\lambda}(\rout)}

\sectionfont{\normalsize}
\subsectionfont{\small}

\begin{document}

\pagestyle{fancy}
\fancyhead{} 
\fancyhead[OR]{\thepage}
\fancyhead[OC]{{\footnotesize{\textsf{DE~SITTER CORES WITH KERR EXTERIORS}}}}
\fancyfoot{} 
\renewcommand\headrulewidth{0.5pt}
\addtolength{\headheight}{2pt} 

\title{{\bf{\Large{\textsf{On a Class of Exact Arbitrarily Differentiable de~Sitter Cores with Kerr Exteriors}}}}\\{\large\textsf{Possible gravastar or regular black hole mimickers}}}

\author {{\small Sa\v{s}a Iliji\'{c} \footnote{sasa.ilijic@fer.hr}} \\
\it{\small Department of Applied Physics,}\\
\it{\small Faculty of Electrical Engineering and Computing, University of Zagreb}\\
\it{\small HR-10000 Zagreb, Unska 3, Croatia}
\\[-0.1cm]
\PRLsepsmall\\[-0.5cm]
\and
{\small Andrew DeBenedictis  \footnote{adebened@sfu.ca}} \\
\it{\small Simon Fraser University} \\
\it{\small Burnaby, British Columbia, V5A 1S6, Canada}\\
\it{\small and}\\
\it{\small The Pacific Institute for the Mathematical Sciences}
}
\date{{\small November 25, 2024}}
\maketitle

\setcounter{footnote}{0}
\begin{abstract}
\noindent Within the paradigm of non-perturbative Einstein gravity we study continuous curvature manifolds which possess de~Sitter interiors and Kerr exteriors. These manifolds could represent the spacetime of rotating gravastars or other similar black hole mimickers. The scheme presented here allows for a $C^{n}$ metric transition from the exactly de~Sitter interior to the exactly Kerr exterior, with $n$ arbitrarily large. Generic properties that such models must possess are discussed, such as the changing of the topology of the ergosphere from $S^{2}$ to $S^{1}\times S^{1}$. It is shown how in the outer layers of the transition region (the ``atmosphere'' as it is often called in astrophysics) the dominant/weak and strong energy conditions can be respected. However, much like in the case of its static spherically symmetric gravastar counterpart, there must be some assumptions imposed in the atmosphere for the energy conditions to hold. These assumptions turn out to not be severe. The class of manifolds presented here are expected to possess all the salient features of the fully generic case. Strictly speaking, a number of the results are also applicable to the locally anti-de~Sitter core scenario, although we focus on the case of a positive cosmological constant.
\end{abstract}
\rule{\linewidth}{0.2mm}
\vspace{-1mm}

\section{{Introduction}}
One of the most fascinating properties of almost any gravitational theory is the prediction of black holes. As is well-known, these objects allow matter to fall into them, but classically do not allow escape. The surface of no return is known as the event horizon. To date, the best theory of gravitation is general relativity, and the vacuum black holes it predicts are of the Kerr family (Kerr-Newman-de~Sitter if the vacuum possesses a cosmological constant and one relaxes the definition of vacuum to include electromagnetic fields) \cite{ref:kerr}, \cite{ref:kerrndes}. It has been shown within general relativity that if a black hole forms and its exterior is essentially vacuum and asymptotically flat, the result must eventually settle down to the Kerr spacetime \cite{ref:hawkell}, \cite{ref:teukbhunique}, although there are some assumptions in this result which have yet to be fully explored \cite{ref:carterbhunique}. The presence of a negative cosmological constant does allow, in spacetimes with non-trivial topology, for other types of black holes (cylindrical, toroidal, and higher genus \cite{ref:topostart}-\cite{ref:topoend}) but we do not consider such exotic cases here. Therefore, in some sense the Kerr family of black holes yields a relatively generic end state for the non-exotic gravitational collapse of massive objects above a certain compactness limit.

The Kerr-de~Sitter spacetime is quite interesting. Its line element, in Boyer-Lindquist coordinates, is furnished by \cite{ref:boylind}, \cite{ref:carterkerr}, \cite{ref:abbassi}
\begin{align}
ds^{2}& = \mathbf{g}[\mathrm{d}\mathbf{x},\mathrm{d}\mathbf{x}] = \left(r^{2}+a^{2}\cos^{2}\theta\right)\left(\frac{{{\rm{d}}}r^{2}}{\Delta}+ \frac{{\rm{d}}\theta^{2}}{1+\frac{\Lambda}{3}a^{2}\cos^{2}\theta}\right) \nonumber \\
& +\sin^{2}\theta\, \frac{1+\frac{\Lambda}{3}a^2\cos^{2}\theta}{r^{2}+a^2\cos^{2}\theta} \left(\frac{a\,{\rm{d}}t -(r^2+a^2)\,{\rm{d}}\phi}{1+ \frac{\Lambda}{3}a^2}\right)^{2} -
\frac{\Delta}{r^{2}+a^{2}\cos^{2}\theta} \left(\frac{{\rm{d}}t -a \sin^{2}\theta\,{\rm{d}}\phi}{1+\frac{\Lambda}{3}a^{2}}\right)^{2}\,,\label{eq:kerrdsmet}
\end{align}
where in this manuscript we restrict the coordinate domain as
\begin{equation}
t_{1} < t < t_{2},\;\; 0 < r_{1} \leq r < \infty,\;\; 0 < \theta < \pi , \;\; 0 \leq \phi < 2\pi\,, \label{eq:coorddomain}
\end{equation}
save for any singularities. Strictly speaking, for geodesic completeness, the structure of the metric should also be specified for $r < r_{1}$, which could in principle transition to something other than de~Sitter spacetime if desired. Here however we restrict our analysis to the domain above. 

In (\ref{eq:kerrdsmet}) we have
\begin{equation}
 \Delta:=r^{2} -2Mr+a^{2}-\frac{\Lambda}{3} r^{2}(r^{2}+a^{2})\,, \nonumber
\end{equation}
with $M$ the mass parameter and $a$ the angular momentum per unit mass ($J=Ma$ with $J$ the angular momentum).
$\Lambda$ is the cosmological constant, assumed positive in this work, although it does not have to be. In the limit $\Lambda \rightarrow 0$ we have the pure Kerr spacetime, of relevance to this paper  as it will provide the exterior geometry. In the limit $M \rightarrow 0$ the metric (\ref{eq:kerrdsmet}) yields locally (anti) de~Sitter spacetime in oblate spheroidal coordinates, also relevant in this paper since it will provide the core's geometry. That this limit is de~Sitter spacetime can be verified from the constancy of the orthonormal Riemann tensor components in this limit, where the constant components only depend on $\Lambda$. This de~Sitter spacetime might possess peculiar topological properties \cite{ref:topodesit}. If $a$ and $\Lambda \rightarrow 0$ in (\ref{eq:kerrdsmet}) then we have the famous Schwarzschild spacetime. On the other hand, if both $M$ and $\Lambda$ are zero with $a\neq 0$ we have Minkowski spacetime in oblate spheroidal coordinates. The topology of the resulting Minkowski space may be non-trivial \cite{ref:topomink}, \cite{ref:topomink2}. A comprehensive review of the Kerr spacetime is provided in \cite{ref:wiltbook}.

\begin{figure}[htbp]
\begin{center}
\includegraphics[width=0.60\textwidth, clip, keepaspectratio=true]{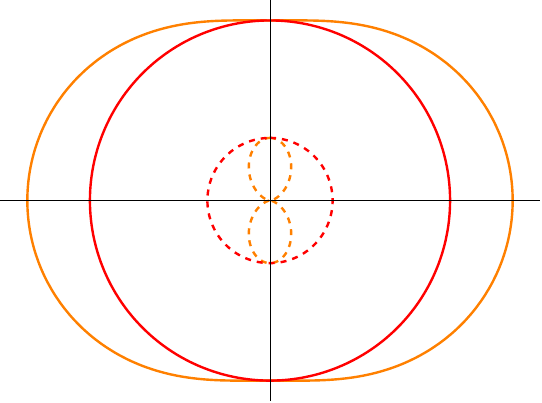}
\caption{{\small{Domains of the pure Kerr spacetime. The outer orange line represents the outer limit of the ergoregion. The solid red line represents the event horizon. The dashed red line represents the inner Cauchy horizon, and the dashed orange line the inner limit of the ergoregion. Here the Boyer-Lindquist $r$ coordinate directly measures distance from the center of the plot, hence the peculiar shape of the inner ergosurface. (i.e. In this plot, and those that follow, the BL $r$ and $\theta$ coordinates are being utilized as ``polar'' coordinates.)}}}
\label{fig:kerrdomains}
\end{center}
\end{figure}

For the moment we shall set $\Lambda=0$ and discuss the pure Kerr spacetime, as this will provide the exterior geometry of the system to be studied. The pure Kerr geometry possesses an event horizon located at
\begin{equation}
r_{\tinyrmsub{H}}:=M+\sqrt{M^2-a^2}\,, \label{eq:kerrhor}
\end{equation}
and a curvature singularity located at $r^{2}+a^{2}\cos^{2}\theta=0$, that is, $r=0$ with $\theta=\pi/2$. This is actually a ``ring'' singularity since in the Boyer-Lindquist coordinates $r=0$ is not a point.  There is a second horizon located at $r=M-\sqrt{M^{2}-a^{2}}$, however it is not strictly speaking an event horizon, but it is a Cauchy horizon. Of particular interest in the Kerr geometry is the presence of the ergoregion, which exists in the domain $M-\sqrt{M^2-a^2\cos^{2}\theta} < r <  M+\sqrt{M^2-a^2\cos^{2}\theta}$. In the ergoregion a timelike particle cannot remain static relative to an observer at infinity. The various domains of the pure Kerr geometry are illustrated in figure \ref{fig:kerrdomains} where in the figure the Boyer-Lindquist coordinate $r$ represents distance from the center of the plot, and $\theta$ altitude angle measured from the vertical. i.e. in the plot $r=0$ is compressed to a point at the center, and this gives rise to the peculiar shape of the inner ergosurface. It is assumed throughout this manuscript that $|a| < M$.

%

An issue that has caused concern in the theoretical study of black holes is the inevitable singularity that resides behind their event horizons. In short, the general theory of relativity predicts that unless energy conditions are violated, the domain within a black hole's event horizon must possess a singularity. (See \cite{ref:singthmthesis} for a nice review.) It is generally believed that the singularity is not physical, but that instead yet unknown high-energy effects will replace the singularity with something more benign. One popular model is that at sufficiently high energies a phase transition occurs \cite{ref:gravastarorig} \cite{ref:gravastarsaibal} in the semi-classical vacuum and at least some region in the interior of a black hole is replaced with a de~Sitter geometry of constant curvature. This is the motivation for the present work here.

As mentioned previously, the de~Sitter spacetime in oblate spheroidal coordinates is given by the metric (\ref{eq:kerrdsmet}) in the vanishing mass limit. That is, the pure de~Sitter line element is provided by
\begin{align}
ds_{\tinyrmsub{dS}}^{2}&= -\frac{3 \left(-\Lambda  \left(a^2+2 r^2\right)+a^2 \Lambda  \cos 2 \theta+6\right)}{2 \left(a^2 \Lambda
+3\right)^2}{\mathrm d}t^{2}  -\frac{6 a \Lambda  \left(a^2+r^2\right) \sin ^2\theta}{\left(a^2 \Lambda +3\right)^2}\, {\mathrm{d}}t\,\mathrm{d}\phi
 \nonumber \\
&+\frac{3 \left(a^2 \cos ^2\theta+r^2\right)}{\left(a^2+r^2\right) \left(3-\Lambda  r^2\right)}\,\mathrm{d}r^{2} +\frac{3 \left(a^2 \cos ^2\theta+r^2\right)}{a^2 \Lambda  \cos ^2\theta+3}\,\mathrm{d}\theta^{2} +  \frac{3 \left(a^2+r^2\right) \sin ^2\theta}{a^2 \Lambda +3}\,\mathrm{d}\phi^{2}\,. \label{eq:desmet}
\end{align}
Although not obvious, the metric yielding line element (\ref{eq:desmet}) does indeed describe a manifold with constant curvature as all orthonormal-frame Riemann tensor components have a constant magnitude. The curvature is positive if $\Lambda >0$ and negative if $\Lambda < 0$. As is well known, the de~Sitter spacetime also possesses a horizon, sometimes referred to as the cosmological horizon, which in this coordinate system is located at
\begin{equation}
 r_{\tinyrmsub{dS}}=\frac{\sqrt{6-2 a^2 \Lambda  \sin^{2} \theta}}{\sqrt{2\Lambda }} \,. \label{eq:deshor}
\end{equation}

The qualitative idea behind de~Sitter core black holes or black hole mimickers is simple. It is now fairly certain that the Kerr geometry accurately describes the geometry outside of physical black hole event horizons \cite{ref:bertiastro}-\cite{ref:EHTkerr}. Therefore one wishes to keep the Kerr geometry outside of the object. However, in the interior, and especially near the high curvature region in the vicinity of the singularity, we lack the guidance of observational evidence. We therefore need to rely on theoretical considerations alone in order to attempt to determine the geometry of the high energy regime.

One proposed alternative for the black hole interior is the aforementioned de~Sitter spacetime. That is, some region of the interior domain of the black hole is replaced with the de~Sitter constant curvature geometry which eliminates the curvature singularity. This idea stems back at least to the work of Gliner where it was proposed that various Einstein spaces could meet the criteria of a physically acceptable vacuum \cite{ref:gliner}. Later, a similar idea of de~Sitter vacua was explored in a cosmological context \cite{ref:starobdesit}, \cite{ref:markovdesit}. Such models were then extended to the Schwarzschild black hole, with a patching at the event horizon, in \cite{ref:shenzhudesit}. Further early analysis, and a review of the state of the field at the time, can be found in the article by Poisson and Israel \cite{ref:poisisdesit}, and a nice treatise on compact objects, including black hole alternatives, may be found in \cite{ref:camenzindbook}.

The gravastar concept takes the above idea one step further. The gravastar (``gravitational vacuum star'') proposes that near the onset of horizon formation, a phase transition occurs which prevents the horizon from forming \cite{ref:gravastarorig}. Instead, a non-null region, which to an observer at infinity very closely resembles an event horizon, is present where the horizon would normally occur. This transition region bridges the exterior region, which mimics that of a black hole, and an interior region, which the phase transition renders as de~Sitter, and thus avoids the black hole singularity. Since the original proposal of the model, a large number of works have been produced studying various possible gravastar properties. (See some representative works \cite{ref:grava1} - \cite{ref:grava17b} and references therein along with a nice review of the subject in \cite{ref:grava18}. We apologize that we cannot cite all relevant works due to their large number.) Some interesting recent calculations regarding dynamical gravastars, such as properties of their surfaces and stability, may be found in  \cite{ref:adlergrava1}, \cite{ref:adlergrava2}. An interesting spherical, stable model has been constructed utilizing two scalar fields in \cite{ref:nojnash}. The study utilized both a thin shell and a smooth transition.

What we wish to accomplish here is to construct a black hole mimicker in the spirit of the above considerations. Specifically the goal here is to provide a class of exact solutions with the following properties:
\begin{itemize}
 \item The geometry is exactly Kerr in the exterior region (motivated by the success of the Kerr model in recent black hole observations of the near horizon region \cite{ref:bertiastro}-\cite{ref:EHTkerr}).
 \item The core, including the singular region, is replaced with the de~Sitter geometry, inspired by the above arguments.
 \item The transition region (the shell) should be of finite extent and sufficiently smooth. Further the joining of the shell to the Kerr manifold, and to the de~Sitter manifold, should be $C^{n}$ for sufficiently large $n$. We define $C^{n}$ here in the most common way as follows:
 \begin{definition}
  A function $f(x)$ is of class $C^{n}$ on $S$ if the derivatives\, $\mathrm{d}^{q}f(x)/\mathrm{d}x^{q}\;\; \forall \;\;q \in \mathbb{Z}^{+}_{0}\leq n\,$ exist and are continuous on an open set $S$ of the real line.
 \end{definition}
 \item The outer layers of the transition region, called the atmosphere, should possess physically acceptable properties, at least as much as is reasonably possible. Here this means that at the very least as many of the common energy conditions as possible should be respected. This is motivated by the fact that in the outer region of the transition layer the physics is relatively ``mild'', meaning the curvature is generally not large and that any material medium present here is generally not believed to be under extreme stresses.
 \item The inner layers of the transition region must, of course, violate energy conditions since the model here is continuous (required for physicality) and the interior de~Sitter spacetime necessarily violates certain energy conditions.
\end{itemize}

The paper is organized as follows: In section \ref{sec:gravastar} we briefly review the traditional gravastar model. This is followed in section \ref{sec:rotgrava} by the construction of the solutions of this paper describing the Kerr black hole mimicker, where a number important properties are studied in various subsections. Finally, some concluding remarks are made.

\section{A brief review of the traditional gravastar models}\label{sec:gravastar}
\begin{figure}[htbp]
\begin{center}
\includegraphics[width=0.55\textwidth, clip, keepaspectratio=true]{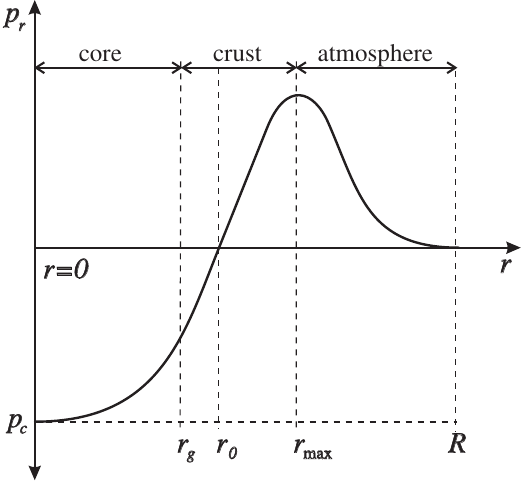}
\caption{{\small{This simplest continuous and differentiable gravastar model with physical atmosphere, taken from \cite{ref:ourgrava} and inspired by \cite{ref:grava1}. The radial pressure ($p_{r}$) is plotted as a function of radius. The de~Sitter negative pressure exists at the center ($p_{c}$), removing the Schwarzschild singularity. In order to respect energy conditions in the atmosphere the radial pressure must transition through a crust region to become positive before falling to zero at the boundary $R$.}}}
\label{fig:gravast-1}
\end{center}
\end{figure}

Since the original proposal of the gravastar \cite{ref:gravastarorig} an enormous amount of work has been done on the subject. The original studies focused on thin-shell models within the paradigm of static spherical symmetry, where the de~Sitter interior was directly patched to the Schwarzschild exterior at some radius \cite{ref:lobode}. In the spirit of mimicking a black hole, this patching was often proposed to occur outside of, but very close to, the Schwarzschild horizon. Soon after, smooth analyses were put forward which eliminated the thin shell and replaced it with a finite transition region. In \cite{ref:ourgrava} it was shown explicitly how gravastars could be made to obey certain continuous and differentiable well-known equations of state, and also satisfy energy conditions in the outer layer. In short, if one wishes the gravastar to smoothly transition between de~Sitter spacetime and Schwarzschild spacetime near, but outside of, the horizon, the gravastar must possess several qualitatively different regions, as illustrated in figure \ref{fig:gravast-1}.

In general, theoretical investigations usually demand that static gravastars possess the following properties:
\begin{itemize}
 \item The deep interior region, at the very least the center, must possess negative and isotropic pressures and positive energy density, due to its de~Sitter nature.
 \item The outer layer, sometimes referred to as the atmosphere, should have pressures approaching zero at the stellar boundary (if one is present). The Synge junction condition in the scenario of static spherical symmetry demands that at least the radial pressure must vanish at the stellar-vacuum boundary.
 \item Physically realistic scenarios require that in the atmosphere the radial pressure should fall to zero at the boundary rather than grow to zero from the negative sector. Similarly, the energy density should be falling within the atmosphere as one proceeds outward from the center. The energy density does not need to be zero at the stellar boundary in the case of static spherical symmetry.
\end{itemize}
The above considerations, along with differentiability, lead to the qualitative picture in figure \ref{fig:gravast-1} taken from \cite{ref:ourgrava}.

It is believed that most astrophysical black holes or black hole mimickers possess at least some amount of angular momentum and recent black hole observations seem to confirm this \cite{ref:rotastro}, \cite{ref:rotastro2}. In the realm of general relativity this means that, at least beyond some distance from the event horizon, the spacetime must be very close to Kerr spacetime. Stationary metrics are much more difficult to deal with than static ones and therefore substantially less work has been performed in the realm of rotating black hole mimickers. Interesting works have been produced in this vein though: The thin-shell formalism has been utilised in \cite{ref:rotgravcardoso}, \cite{ref:rotgrav1} and \cite{ref:rotgrav2} under the assumption of slow rotation. A different slowly rotating model was considered in \cite{ref:rotgrav3}.  In \cite{ref:rotgravX} models have been discussed in the slowly rotating regime up to second-order in the rotation parameter beyond the Schwarzschild case. A further analysis at one-higher order in rotation was performed in \cite{ref:rotgravY}, and possible astrophysical considerations were considered in \cite{ref:astrorotgrav}. Below we create a class of exact and smooth (meaning arbitrary magnitude of rotation up to $|a| =M$, and arbitrarily high differentiability class) geometries of a gravastar-like black hole mimicker.

\section{A class of arbitrarily smooth Kerr black hole mimickers with de~Sitter cores}\label{sec:rotgrava}
Probably the simplest method to connect two otherwise incompatible spacetimes together is via the use of infinitely thin-shells. Although not very physical, the thin-shell formalism does often allow one to glean the interesting physical properties of a transition region by studying the properties of the thin shell itself. However, in a physical theory governed by second-order differential equations, a sufficiently differentiable manifold is generally more desirable, and therefore a $C^{n}$ solution, for sufficiently high $n$, is sought for physical reasons.

\begin{figure}[htbp]
\begin{center}
\includegraphics[width=0.55\textwidth, clip, keepaspectratio=true]{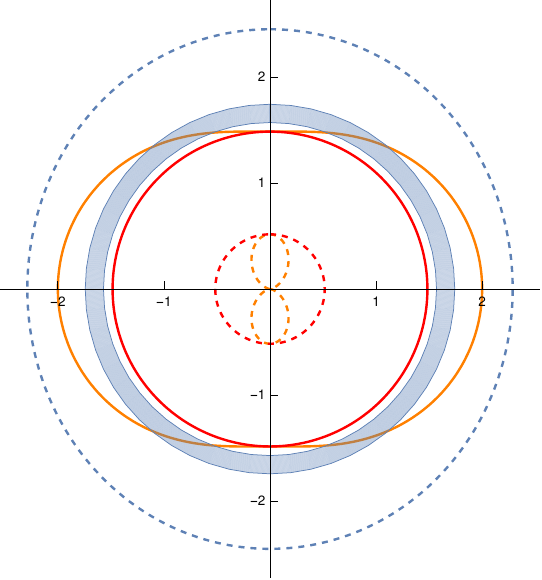}
\caption{{\small{A superposition of the interesting surfaces of de~Sitter spacetime, Kerr spacetime, and the shell. The solid and dashed red and orange lines represent the same Kerr surfaces as in figure \ref{fig:kerrdomains}. The thick blue circle represents the transition shell, and the dashed blue line represent the pure de~Sitter horizon. The outer layer of the shell should be located just outside the Kerr horizon, but inside the de~Sitter horizon, for a black hole mimicker such as a gravastar.}}}
\label{fig:rot_grav_1}
\end{center}
\end{figure}

\begin{figure}[htbp]
\begin{flushright}
\includegraphics[width=0.80\textwidth, clip, keepaspectratio=true]{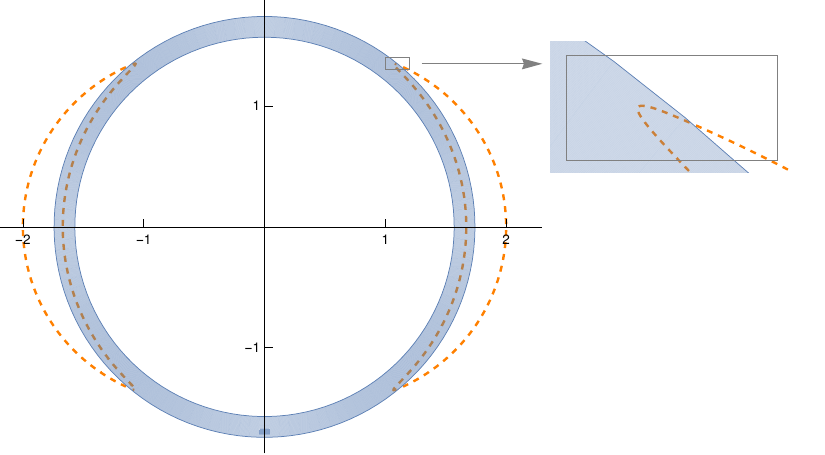}
\end{flushright}
\begin{center}
\caption{{\small{A representative model of the black hole mimicker generated using actual parameters of the models in the paper, which will be discussed below in subsection \ref{sec:toy}. The transition shell is represented as the thick blue region, and the ergosurface is indicated by the dashed orange line. Notice that now the ergosurface has $S^{1}\times S^{1}$ topology. Inset: A close up of where the ergosurface penetrates the transition shell.}}}
\label{fig:rot_grav_2}
\end{center}
\end{figure}

In this section we aim to develop a method which yields a class of continuous compact objects with de~Sitter cores and Kerr exteriors. The core and the exterior will be connected by a transition region, which we refer to as the ``shell''. In order to be compatible with the gravastar model the outer surface of the shell should be placed just outside of the corresponding Kerr black hole's event horizon (\ref{eq:kerrhor}), and we shall denote it as $\rout$. This means that, in the spirit of gravastars, one has an object that is not necessarily a black hole, but mimics one up to arbitrarily close to the corresponding black hole's event horizon. In principle the inner surface of the shell, $\rin$, could be located at any $r_{1} < \rin < \rout$. The qualitative scenario is illustrated in figure \ref{fig:rot_grav_1}, where for illustrative purposes we have placed the entire transition shell just outside the horizon. In the figure the color of the lines indicate the same surfaces as in figure \ref{fig:kerrdomains}, with the addition of two new features: the shell (light blue thick region) and the pure de~Sitter cosmological horizon (outermost dashed blue line). In the models we present here everything inside of the shell will be pure de~Sitter, and everything outside of the shell will be pure Kerr. In other words, most of the lines in figure \ref{fig:rot_grav_1} no longer exist and the model actually is described by figure \ref{fig:rot_grav_2}. In figure \ref{fig:rot_grav_2} we have de~Sitter inside the shell, and Kerr outside the shell and the ergosurface (modified by the presence of the shell) is also illustrated as an orange dashed line. This figure is not simply qualitative but was generated using the actual models proposed below in this manuscript. Of particular interest is how the presence of the shell modifies the ergosurface. In the pure Kerr geometry the ergoregion is bounded by two $S^{2}$ surfaces. However, in the black hole mimicker the ergosurface changes topology to $S^{1}\times S^{1}$. The shell could of course also be located outside of the ergoregion, thus eliminating the vacuum ergoregion altogether.

The main goal here is to construct a manifold with a de~Sitter core and Kerr exterior and which is differentiable everywhere at some arbitrarily high differentiability class $C^{n}$. In the theory of general relativity it is often desirable to have a manifold with $n \geq 2$, although certain jumps in the second derivative of the metric are technically allowed \cite{ref:syngebook}, and at the level of the first derivatives it is the extrinsic curvature which one usually demands the continuity of \cite{ref:senjunct} - \cite{ref:isjunct}. If one further demands a continuous conservation law (strictly not required) then $n \geq 3$ should generally be chosen. The mathematical construction in this manuscript is valid for any $n > 0$, although we shall frame the discussions for $n\geq 3$.

Since the de~Sitter spacetime and the Kerr spacetime (not including the singularities, which do not exist in figure \ref{fig:rot_grav_2}) are already sufficiently smooth, we need to ensure that the boundaries of the transition shell, when joining to de~Sitter and Kerr, meet the smoothness requirement since the job of the shell is to transition metric (\ref{eq:desmet}) to the pure Kerr metric (metric (\ref{eq:kerrdsmet}) with $\Lambda = 0$). Let us facilitate this transition with the following replacements in (\ref{eq:kerrdsmet}):
\begin{align}
 & M=0   &\mbox{for \quad $0 < r_{1} < r < \rin$}\,, &\qquad \mbox{(de~Sitter region)}\,, \nonumber \\ 
 & M\to \mu(r)M  \;\; \mbox{and} \;\; \Lambda \to \lambda(r) \Lambda &\mbox{for}\quad \rin \leq r \leq \rout \,, &\qquad \mbox{(transition shell region)}\,, \label{eq:mlambdafuncs} \\
 & \Lambda= 0 & \mbox{for \quad $r > \rout$}\,, &\qquad \mbox{(Kerr region)}\,, \nonumber
\end{align}
where the functions $\mu(r)$ and $\lambda(r)$ need only be defined for the shell region given by $\rin \leq r \leq \rout$. Further, the function $\mu(r)$ is required to possess the following properties: It should vanish at $\rin$ and equal $1$ at $\rout$. In between these two radial values it can have any properties one wishes, save for its first few derivatives not being infinite. Further, at the end points, $\rin$ and $\rout$, the derivatives of $\mu(r)$ up to some order $n \geq 3$ should be zero, to allow for $C^{n}$ continuity throughout the entire manifold. On the other hand, the other function, $\lambda(r)$, should have a value of $1$ at $\rin$ and a value of zero at $\rout$. Its derivatives at $\rin$ and $\rout$ should also vanish to some order $n \geq 3$ to allow for sufficient smoothness at the patching hypersurfaces. As with $\mu(r)$, $\lambda(r)$ can possess any reasonable properties desired in  between the $\rin$ and $\rout$ that are allowable by the aforementioned boundary conditions. Also, not all $\mu(r)$ and $\lambda(r)$ will eliminate the horizon. To put it loosely, if the spacetime in the horizon region is ``too Kerr-like'', the shell will not be sufficiently ``strong'' to eliminate the horizon. This is not something which is particular to this model. Even in the well-studied static gravastars, the presence of a bridging region between Schwarzschild and de~Sitter is not sufficient to guarantee the absence of a horizon. The gravitational influence of the bridging region in the vicinity of the horizon needs to be strong enough to eliminate it. We will also see below that further restrictions on $\mu(r)$ and $\lambda(r)$ must be imposed near $\rout$ in order to have a physically well behaved atmosphere region. We mention before proceeding that only in the very outer layer of the shell can the quantity $\mu(r)M$ have any approximate interpretation as a mass.

As discussed above the functions $\mu(r)$ and $\lambda(r)$ are to be $C^{n}$ with $n \geq 3$. The $r$-only dependence is motivated by the fact that, in the traditional gravastar model, the phase transition occurs at the onset of horizon formation, and in the coordinates used here the Kerr event horizon is located at constant $r$, with no $\theta$ dependence. We assume therefore that the phase transitioning shell possesses this property as well. One could in principle relax the assumption to functions which depend also on the other coordinates, but unfortunately this comes with the baggage of great technical difficulty in the resulting calculations.

\begin{figure}[htbp]
\begin{center}
\includegraphics[width=0.40\textwidth, clip, keepaspectratio=true]{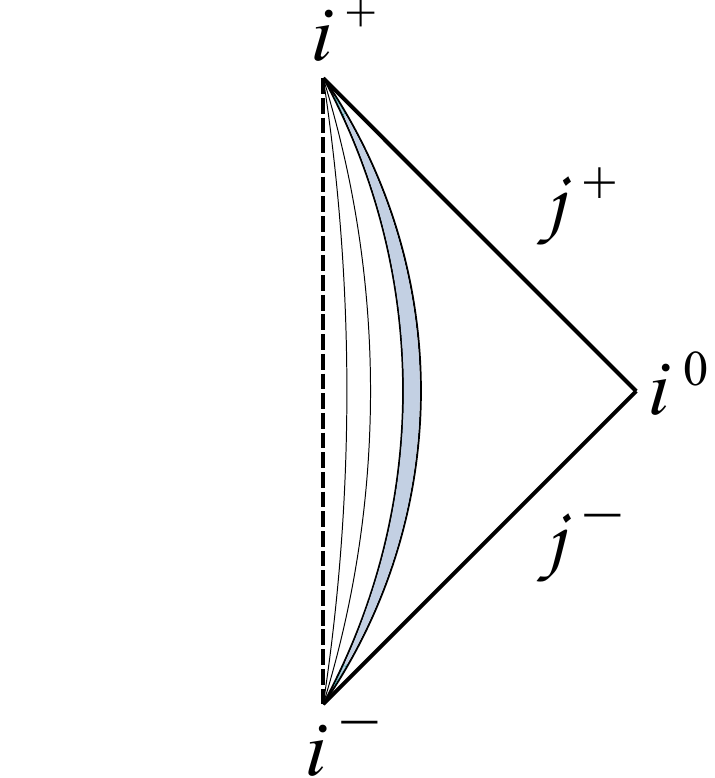}
 \caption{\small A (partial) possible Penrose-Carter diagram for the spacetime under study. The blue region represents the shell which is bounded by de~Sitter on the left and Kerr on the right. The exact structure inside the shell is quite complex and therefore that region may not be accurately depicted. The full structure is not depicted as the spacetime for $r < r_{1}$ is not considered in our domain. (See
(\ref{eq:coorddomain}) and the comments that follow it.)}
\label{fig:PC_diag}
\end{center}
\end{figure}

That such functions $\mu(r)$ and $\lambda(r)$ exist may be proven via the Stone-Weierstrass theorem. In some of the analysis below we will further assume that these functions are analytic ($C^{\omega}$) in the vicinity of $\rout$. We stress here that the behaviour of $\mu(r)$ and $\lambda(r)$ in the domains $r_{1} < r < \rin$ and $r > \rout$ is irrelevant since we reiterate that for $r<\rin$ we employ the pure de~Sitter metric, and for $r > \rout$ the pure Kerr geometry. The structure of the spacetime in the domain (\ref{eq:coorddomain}) is illustrated in figure \ref{fig:PC_diag}. The shaded shell is bounded by de~Sitter on the left and Kerr on the right and the cosmological horizon is eliminated due to the placement of the shell. We caution that the figure is only a highly qualitative representation of the scenario, since the shell region is extremely complicated and therefore the full structure there is in general difficult to discern. We do know however that the structure there is Lorentzian (to be clarified shortly) and regular (also to be clarified below). As mentioned after (\ref{eq:coorddomain}), the structure for $r$ values where $r < r_{1} < \rin$, which may or may not be de~Sitter, is not considered here. (Recall that we are locally patching to de~Sitter spacetime here. What happens further in the interior could remain de~Sitter or be something else depending on the physics one wishes to impose there.)

We also list for convenience the metric inverse, which in simplified notation is given by:
\begin{equation}
 [g^{\mu\nu}]=\left[\begin{array}{cccc}
                     \frac{g_{\phi\phi}(r,\theta)}{g_{\phi\phi}(r,\theta)\,g_{tt}(r,\theta) - g_{t\phi}^{2}(r,\theta)} & 0 & 0 & -\frac{g_{t\phi}(r,\theta)}{g_{\phi\phi}(r,\theta)\,g_{tt}(r,\theta) -g_{t\phi}^{2}(r,\theta)} \\
                     0 & \frac{1}{g_{rr}(r,\theta)} & 0 & 0 \\
                     0 & 0 & \frac{1}{g_{\theta\theta}(r,\theta)} & 0\\
                     -\frac{g_{t\phi}(r,\theta)}{g_{\phi\phi}(r,\theta)\,g_{tt}(r,\theta) -g_{t\phi}^{2}(r,\theta)} & 0 & 0 & \frac{g_{tt}(r,\theta)}{g_{\phi\phi}(r,\theta)\,g_{tt}(r,\theta) -g_{t\phi}^{2}(r,\theta)}
                    \end{array}\right]\,, \label{eq:invmet}
\end{equation}
from which we can see that potential issues might occur if $g_{rr}=0$ or $g_{\phi\phi}\,g_{tt} -g_{t\phi}^{2} =0$.

Since we are bridging de~Sitter spacetime to the Kerr spacetime, which involves an ergosurface, there may be a number of sign changes in the various metric components throughout the manifold. Physically we would prefer that the spacetime remain of Lorentzian signature (+2 in this manuscript). The most straight-forward way to determine the signature of the spacetime is to compute the eigenvalues of the metric tensor, $\mathbf{g}$, \cite{ref:dasdebbook}. If the spacetime is everywhere Lorentzian this should yield one negative eigenvalue and three positive eigenvalues at every point in spacetime. That is, we calculate all $\ell$ that satisfy\footnote{Please see Appendix I for a brief discussion of this formula; specifically what is, and is not, useful about it.}
\begin{equation}
 \det\left[\mathbf{g}-\ell \mathbf{1}\right]=0\,, \label{eq:evaleqn}
\end{equation}
and by Sylvester's law of inertia the number of positive, negative (and zero) eigenvalues will be basis independent. The characteristic polynomial (\ref{eq:evaleqn}) admits the following four solutions:
\begin{subequations}
\begin{align}
 \ell_{(1)}&=g_{\theta\theta}\,, \label{eq:eval1}\\
 \ell_{(2)}&=g_{rr}\,, \label{eq:eval2}\\
 \ell_{(3)}&=\frac{1}{2}\left[g_{tt}+g_{\phi\phi} + \sqrt{\left(g_{tt}-g_{\phi\phi}\right)^{2}+4 g_{t\phi}^{2}}\,\right]\,,\label{eq:eval3} \\
 \ell_{(4)}&=\frac{1}{2}\left[g_{tt}+g_{\phi\phi} - \sqrt{\left(g_{tt}-g_{\phi\phi}\right)^{2}+4 g_{t\phi}^{2}}\,\right]\,.\label{eq:eval4}
\end{align}
\end{subequations}
The explicit form of each eigenvalue is tabulated in Appendix I.  At this stage we can state the following regarding the eigenvalues (\ref{eq:Aeval1}) - (\ref{eq:Aeval34}):
\begin{itemize}
 \item The eigenvalue $\ell_{(1)}$ is positive provided the function $\lambda(r)$ obeys the following inequality, assuming $a \neq 0$:
 \begin{equation}
  \lambda(r) \Lambda > -\frac{3}{a^{2}\cos^{2}\theta}\,, \label{eq:lamdarestriction}
 \end{equation}
and of course the eigenvalue is negative if the inequality is flipped.
\item The eigenvalue $\ell_{(2)}$ is positive if the function $\mu(r)$ obeys this inequality for $r>0$:
\begin{equation}
 \mu(r) M < \frac{\left(3-r^{2}\lambda(r)\Lambda\right)(a^{2}+r^{2})}{6r}\,. \label{eq:murestriction}
\end{equation}
If the inequality (\ref{eq:murestriction}) is reversed, the eigenvalue is negative.
In principle $\mu(r)$ could be negative in the inner regions of the shell, since energy conditions must be violated there, and also $\mu(r) M$ is not a direct measure of mass there.
\item The other two eigenvalues, $\ell_{(3)}$ and $\ell_{(4)}$, are rather complicated and hence need to be evaluated with care in specific models. We will show below that it is easily possible to obtain models with signature +2 everywhere.
\end{itemize}

Because of the complication introduced by the eigenvalues $\ell_{(3)}$ and $\ell_{(4)}$ we introduce as a secondary, though slightly less robust, test of Lorentzian structure provided by the determinant of the metric. The determinant should be of one sign if there is no signature change in the shell \footnote{Strictly speaking, in four dimensions the metric's signature could change from $+2$ to $-2$ and its determinant would not change sign. Hence the determinant is not as robust of a test of signature change as a direct study of the metric eigenvalues.}. The determinant of the metric turns out to be quite simple as
\begin{equation}
 \mbox{det}\left[\mathbf{g}\right]=-\frac{81\left(a^{2}\cos^{2}\theta +r^{2}\right)^{2}\sin^{2}\theta}{\left(\Lambda \lambda(r)a^{2} +3 \right)^{4}} \,,  \label{eq:detg}
\end{equation}
and we note that in the shell region this determinant nowhere changes sign for real values of the parameters.

Calculating the Einstein tensor, $\mathbf{G}$, with metric (\ref{eq:kerrdsmet}) where $M$ and $\Lambda$ have been replaced via the prescription (\ref{eq:mlambdafuncs}), will give us information regarding the stress-energy content of the transition shell via the Einstein equations
\begin{equation}
 \mathbf{G}=8\pi \mathbf{T}\,. \label{eq:einsteqn}
\end{equation}
Here $\mathbf{T}$ is of course the stress-energy tensor, which in this work is defined from the Einstein tensor via (\ref{eq:einsteqn}) and includes any effective cosmological constant. This stress-energy, subject to prescription (\ref{eq:mlambdafuncs}), turns out to possess the following generic structure in the coordinate frame (\ref{eq:kerrdsmet})
\begin{equation}
\mathbf{T}:= \left[T^{\mu}_{\;\;\nu}\right]= \left[\begin{array}{cccc}
                                      T^{0}_{\;\;0} & 0 & 0 & T^{0}_{\;\;3} \\
                                      0 & T^{1}_{\;\;1} & T^{1}_{\;\;2} & 0 \\
                                      0 & T^{2}_{\;\;1} & T^{2}_{\;\;2} & 0 \\
                                      T^{3}_{\;\;0} & 0 & 0 & T^{3}_{\;\;3}
                                     \end{array}\right]\,. \label{eq:Tudgeneric}
\end{equation}
That is, aside from the energy density and principal pressures ($T^{0}_{\;\;0}$ and $T^{i}_{\;\;i}$ respectively), the shell generally possesses a shear ($T^{1}_{\;\;2}$ and $T^{2}_{\;\;1}$) as well as an azimuthal energy flux ($T^{0}_{\;\;3}$ and $T^{3}_{\;\;0}$). The azimuthal flux is not unexpected since the material making up the shell must be rotating about the symmetry axis. Due to the previously mentioned properties of $\mu(r)$ and $\lambda(r)$, at $r=r_{\tinyrmsub{out}}$ all components of $\mathbf{T}$ vanish, and at $r=r_{\tinyrmsub{in}}$ the structure automatically becomes
\begin{equation}
T^{\mu}_{\;\;\nu}(r_{\tinyrmsub{in}})= -\frac{\Lambda}{8\pi} \delta^{\mu}_{\;\;\nu}\,, \label{eq:TuddeSitter}
\end{equation}
as required for the de~Sitter spacetime.

We should also comment on the extrinsic curvature here, as its continuity in general relativity is required for the absence of ``delta function'' shells. Of particular interest to us is the extrinsic curvature's behavior in the vicinity of the patching hypersurfaces, which are given by the level surfaces $r-\rin=0$ and $r-\rout=0$. We consider the radially outward pointing unit normal, $\hat{n}^{\mu}_{(r)}$, in the neighborhood of the hypersurfaces. The extended extrinsic curvature is then given by the following Lie derivative
\begin{equation}
 K_{\mu\nu}=\frac{1}{2}\mathcal{L}_{\hat{n}^{\mu}_{(r)}}\left[g_{\mu\nu}-\varepsilon\hat{n}_{\mu}^{(r)}\hat{n}_{\nu}^{(r)}\right]\,, \label{eq:extcurv}
\end{equation}
with $\varepsilon:= \hat{n}^{\!(r)\,\alpha}\hat{n}_{\alpha}^{(r)}$. Expression (\ref{eq:extcurv})  yields
\begin{equation}
 [K_{\mu\nu}]=\frac{1}{2}\left[\begin{array}{cccc}
                          \frac{\partial_{r}g_{tt}(r,\theta)}{\sqrt{g_{rr}(r,\theta)}} & 0 & 0 & \frac{\partial_{r}g_{t\phi}(r,\theta)}{\sqrt{g_{rr}(r,\theta)}}\\
                          0 & 0 & 0 & 0 \\
                          0 & 0 & \frac{\partial_{r}g_{\theta\theta}(r,\theta)}{\sqrt{g_{rr}(r,\theta)}} \\
                          \frac{\partial_{r}g_{t\phi}(r,\theta)}{\sqrt{g_{rr}(r,\theta)}} & 0 & 0 & \frac{\partial_{r}g_{\phi\phi}(r,\theta)}{\sqrt{g_{rr}(r,\theta)}}
                         \end{array}\right]\,. \label{extcurvmatrix}
\end{equation}
Ignoring the $r^{\mbox{\footnotesize th}}$ row and column gives the usual extrinsic curvature for the hypersurface considered here. We note from the above expression that continuity of the metric, along with its first derivatives, will ensure continuity of the extrinsic curvature at $\rin$ and $\rout$. We should avoid $g_{rr}(r,\theta)=0$.

At this stage it is useful to introduce an orthonormal tetrad, $[e^{\hat{\alpha}}_{\;\;\mu}]$, in order to project relevant tensors into a physical orthonormal frame. We will denote orthonormal frame indices here with a hat. There are of course infinitely many orthonormal frames one may choose. It is convenient to choose one of the form
\begin{equation}
 \left[e^{\hat{\alpha}}_{\;\;\mu}\right]=\left[\begin{array}{cccc}
                                              \frac{\sqrt{g_{t\phi}^{2}-g_{tt}g_{\phi\phi}}}{\sqrt{g_{\phi\phi}}} & 0 & 0 & 0\\
                                              0 & \sqrt{g_{rr}} & 0 & 0 \\
                                              0 & 0 & \sqrt{g_{\theta\theta}} & 0\\
                                              \frac{g_{t\phi}}{\sqrt{g_{\phi\phi}}}& 0 & 0 & \sqrt{g_{\phi\phi}}
                                             \end{array}\right]\,. \label{eq:hUd}
\end{equation}
It can be verified by straight-forward computation that
\begin{equation}
 e^{\hat{\alpha}}_{\;\;\mu}e^{\hat{\beta}}_{\;\;\nu}\eta_{\hat{\alpha}\hat{\beta}}=g_{\mu\nu}\qquad \mbox{and} \qquad   e^{\hat{\alpha}}_{\;\;\mu}e^{\hat{\beta}}_{\;\;\nu}g^{\mu\nu}= \eta^{\hat{\alpha}\hat{\beta}}\,, \label{eq:hUdmetcompatible}
\end{equation}
with $[\eta^{\hat{\alpha}\hat{\beta}}]=\mbox{diag}[-1,\,1,\,1,\,1]$ and $g_{\mu\nu}$ the metric (\ref{eq:kerrdsmet}) with the substitution (\ref{eq:mlambdafuncs}).
The tetrad (\ref{eq:hUd}) shall be sufficient for now, though we need to be careful with the sign in the radical. Of specific importance is that the vector $\mathbf{e}^{\hat{0}}$ is everywhere timelike with respect to the inner-product of the spacetime metric $[g^{\mu\nu}]$ and the $\mathbf{e}^{\hat{i}}$ are spacelike.

The orthonormal Riemann tensor components,
\begin{equation}
R_{\hat{\alpha}\hat{\beta}\hat{\gamma}\hat{\delta}} = R_{\mu\nu\rho\sigma}e_{\hat{\alpha}}^{\;\;\mu} \, e_{\hat{\beta}}^{\;\;\nu} \, e_{\hat{\gamma}}^{\;\;\rho} \,  e_{\hat{\delta}}^{\;\;\sigma}\, , \label{eq:orthriem}
\end{equation}
whose values if infinite herald a curvature singularity, are rather complicated so not very perspicuous and we do not list them here. We do make the following comments about their form though: If the functions $\mu(r)$ and $\lambda(r)$ are twice differentiable (which they are) then some of the orthonormal Riemann components in the transition shell could \emph{potentially} blow up if $(g_{t\phi})^{2} = g_{tt}\,g_{\phi\phi}$ or if $g_{rr}$, $g_{\theta\theta}$ or $g_{\phi\phi}$ vanish. We stress that these are potential singularities only, meaning that if in the transition shell these conditions can all be avoided, we for certain do not possess a singularity in the orthonormal Riemann components. On the other hand, if some of these conditions are not avoided, we \emph{might} have a singularity. As an example, at the poles ($\theta=0,\, \pi$) $g_{\phi\phi}$ vanishes, yet there is generally no curvature singularity there and the orthonormal Riemann components are all finite.

Before continuing we mention the important, but here neglected, issue of stability. Even though to our knowledge this is the first exact $C^{n}$ gravastar-like solution with Kerr exterior, there are studies in the literature which imply that such objects may have an instability issue \cite{ref:rotgravstab1}. It has also been argued that the instability may not be universal \cite{ref:rotgravstab2}, however, at sufficiently large values of angular momentum it might re-emerge \cite{ref:rotgravstab3}. The way the class of models here is constructed essentially forces the polar moments in the phase transition shell to be compatible with the Kerr exterior and this balance could be unstable.

\subsection{A simple ``toy'' model} \label{sec:toy}
We discuss briefly here a specific and rather simple model. This specific model does not attempt to create a realistic atmosphere but does serve to concretely illustrate all the other salient features expected in such a black hole mimicker such as continuity, regularity, and Lorentzian structure.

We wish to build this simple model of a shell
that renders all metric components continuous
up to some sufficiently high-order of the radial derivative within the shell. We also wish a $C^{n}$ patching at
the inner hypersurface
where the shell joins the de~Sitter interior ($r=\rin$), as well as at the outer hypersurface  ($r=\rout$)
where it joins the Kerr exterior spacetime. This will be done according to the above procedure. 
We make use of a polynomial model of a distributed step function
transitioning from $0$ at the inner hypersurface to $1$ at the outer hypersurface.
We first write $p(x) = \sum_{k=0}^{2n+1} c_k x^k$,
where we require $p(-1) = 0$, $p(1) = 1$,
and  $p^{(m)}(\pm1) = 0$ for $m=1,\ldots,n$, ($p^{(m)}$ being the $m$-th derivative of $p$)
which makes all derivatives up to $n$-th order
vanish at both ends of the transition domain.
These conditions of vanishing derivative allow for the straightforward computation
the $2n+2$ required coefficients $c_k$, $k=0,\ldots,2n+1$.
The functions $\mu(r)$ and $\lambda(r)$
introduced earlier are now constructed in this simple polynomial model as
\begin{equation}
  \mu(r) = 1 - \lambda(r) =
p\big( (r - (r_{\mathrm{in}} + r_{\mathrm{out}})/2)
     / ((r_{\mathrm{out}} - r_{\mathrm{in}})/2) \big) ,
\qquad (r_{\mathrm{in}} \le r \le r_{\mathrm{out}})\,, \label{eq:polystep}
\end{equation}
where we again stress that this very specific form is not general, but only for illustrative purposes of this  simple example.
 
\begin{figure}[htbp]
\begin{center}
\includegraphics[width=0.60\textwidth, clip, keepaspectratio=true]{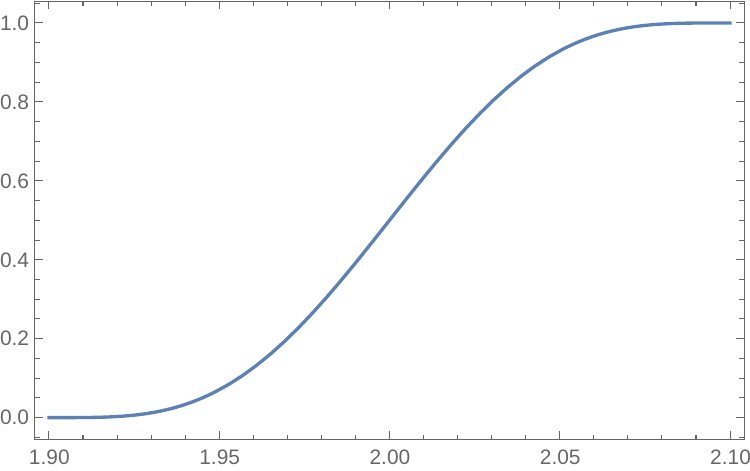}
\caption{{\small{A simple, sample function for $\mu(r)$ and $1-\lambda(r)$ given by the polynomial (\ref{eq:polycoeffs}). At the end points derivatives up to order $n=3$ vanish.}}}
\label{fig:polyfun}
\end{center}
\end{figure}

To explicitly show the results we set the continuity level of the patching here to $n=3$, so that here our polynomial function is specifically
\begin{equation}
 p(x)=\frac{1}{2} + \frac{35}{32}x -\frac{35}{32}x^{3}+ \frac{21}{32}x^{5} -\frac{5}{32} x^{7}\,, \label{eq:polycoeffs}
\end{equation}
where $x$, as in (\ref{eq:polystep}), is defined as $x:= (r - (r_{\mathrm{in}} + r_{\mathrm{out}})/2)/ ((r_{\mathrm{out}} - r_{\mathrm{in}})/2)$. The coefficients in (\ref{eq:polycoeffs}) are unique since, as discussed above, they are set by requiring that the function has vanishing derivatives up to order $n$ at $\rin$ and $\rout$. We display this polynomial function in figure \ref{fig:polyfun}.

\begin{figure}[htbp]
\begin{center}
\includegraphics[width=0.950\textwidth, clip, keepaspectratio=true]{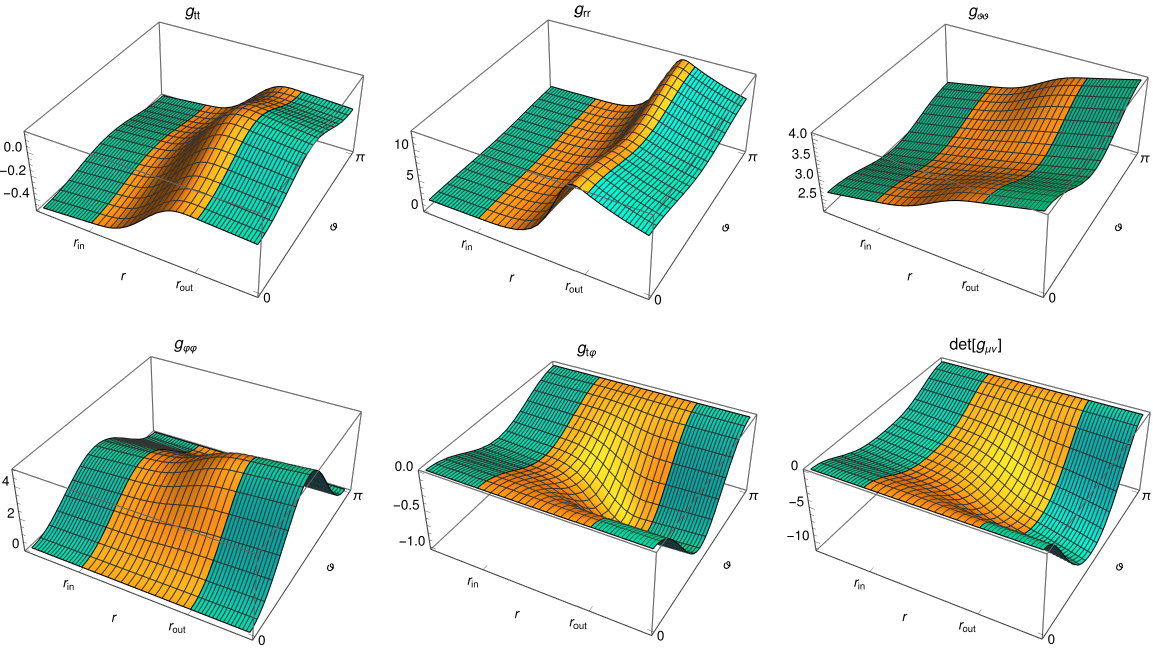}
\caption{{\small{The metric functions for the simple monotonically increasing polynomial model. The yellow region represents the transition shell, and the left and right green regions are parts of the de~Sitter and Kerr spacetimes respectively.}}}
\label{fig:polymetric}
\end{center}
\end{figure}

\begin{figure}[htbp]
\begin{center}
\includegraphics[width=1.00\textwidth, clip, keepaspectratio=true]{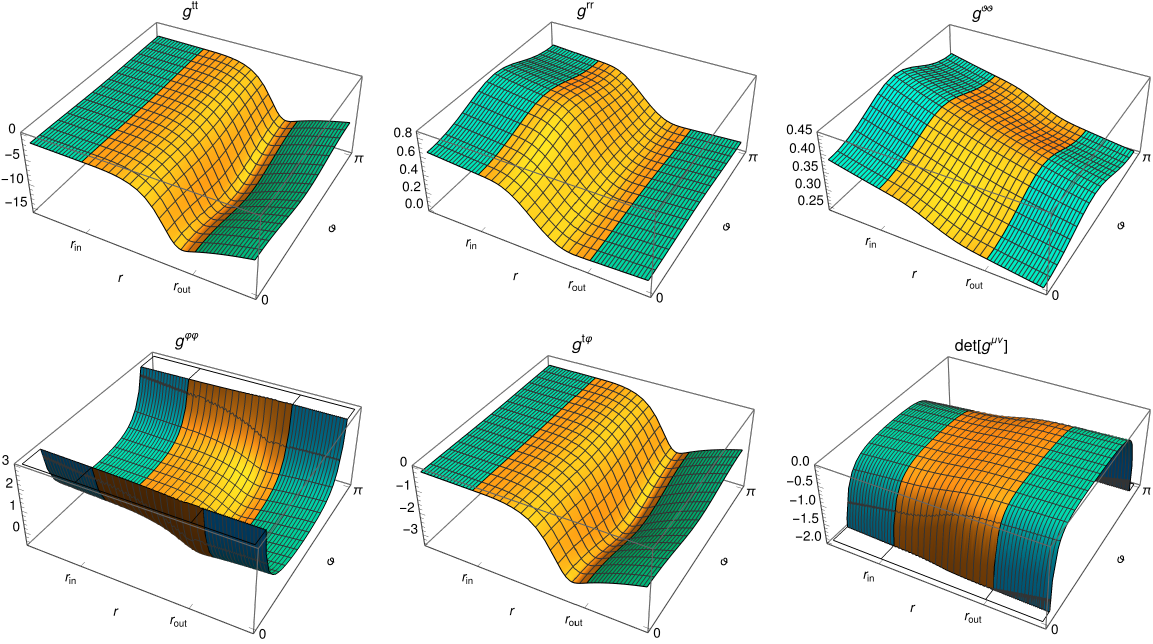}
\caption{{\small{The components of the inverse metric functions for the simple monotonically increasing polynomial model. The yellow region represents the transition shell, and the left and right green regions are parts of the de~Sitter and Kerr spacetimes respectively. The inverse metric components are well behaved save for the standard coordinate issue at the poles, which have been excised in several of the plots in the figure.}}}
\label{fig:polyinmet}
\end{center}
\end{figure}

Using the above polynomial in the metric we can plot the metric components from $r<\rin$ to $r > \rout$ and $0\leq\theta\leq\pi$. These are illustrated in figure \ref{fig:polymetric} along with the determinant of the metric over the same region. In this sub-section we use $M=1$, $a=7/8$, and $\Lambda=1/2$, which gives rise to figure \ref{fig:rot_grav_2}, but the general qualitative features presented here do not depend on the specific values of the parameters provided $M > |a|$. We also illustrated that none of the inverse metric components blow up in the domain of consideration, save for the the usual coordinate issue at the poles, in figure \ref{fig:polyinmet}.

\begin{figure}[H]
\begin{center}
\includegraphics[width=1.00\textwidth, clip, keepaspectratio=true]{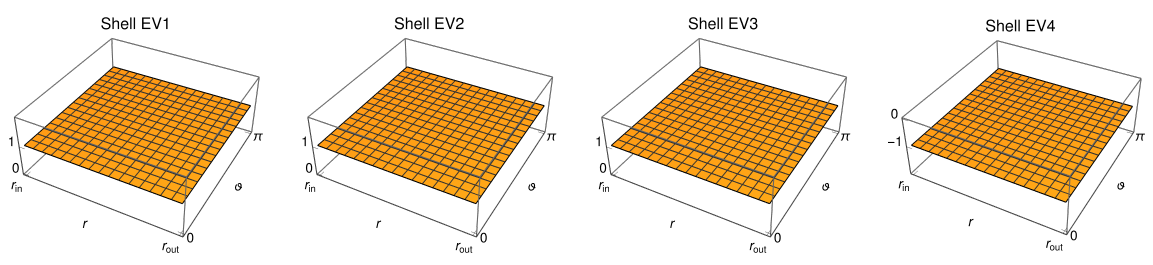}
\caption{{\small{The (usual) eigenvalues of the metric divided by their respective magnitudes. Note that there is one negative eigenvalue and three positive eigenvalues everywhere, indicating an everywhere Lorentzian spacetime of signature +2.}}}
\label{fig:polyevals}
\end{center}
\end{figure}

We next illustrate the Lorentzian structure of the spacetime by plotting the eigenvalues (\ref{eq:eval1})-(\ref{eq:eval4}) divided by their respective magnitudes. We show these in the plots in figure \ref{fig:polyevals} where it may be seen that throughout the entire shell (and therefore throughout the entire spacetime) one eigenvalue is negative and three are positive.

\begin{figure}[htbp]
\begin{center}
\includegraphics[width=0.485\textwidth, clip, keepaspectratio=true]{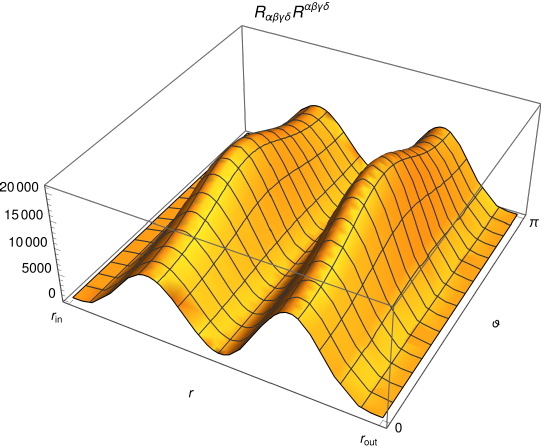}
\caption{{\small{The Kretschmann scalar, $R_{\mu\nu\sigma\rho}R^{\mu\nu\sigma\rho}$, in the transition layer. The Kretschmann scalar is finite everywhere within the transition layer.}}}
\label{fig:polyriem}
\end{center}
\end{figure}

Finally, in order to facilitate illustrating that the spacetime is regular we plot the Kretschmann scalar, $K:=R_{\mu\nu\sigma\rho}R^{\mu\nu\sigma\rho}$ in figure \ref{fig:polyriem}. We note that $K$ is finite everywhere within the transition region. 

\subsection{Analysis of a physical atmosphere}
The above model was rather simple, and certainly no attempt was made to model reasonable physics in the outer layer. However it did serve to illustrate how the scheme is capable of creating a gravastar-like black hole mimicker with de~Sitter interior, Kerr exterior, and a sufficiently differentiable spacetime throughout, which is our main goal. In the gravastar literature it is generally desired, from a physical perspective, that the outer layer of the gravastar possess reasonable physics, or at least as reasonable as possible, due to the non-extremality of physical conditions there. We here therefore concentrate on the outer region of the transition shell in a more realistic model. For this we note that for a continuity level $n$ patching to the Kerr metric at $r=\rout$, the metric function $\mu(\rout)$ must equal $1$ and its first $n$ derivatives must vanish at $r=\rout$. We assume that $\mu(r)$ is $C^{\omega}$ (real analytic) in some non-zero domain $(\rout-\varepsilon) < r \leq \rout$ (which could in principle be the entire domain of the transition shell) and write this function in this domain as \footnote{We stress here for clarity that even if the functions $\mu(r)$ and $\lambda(r)$ are analytic, it does \emph{not} mean that they match the Kerr metric at $r=\rout$ analytically. The shell functions may be $C^{\omega}$, but their matching to the Kerr metric at $\rout$ is only $C^{n}$, although $n$ can be arbitrarily large.}
\begin{equation}
 \mu(r)=1-\left(\rout-r\right)^{n+1}\gsm\,, \label{eq:gsm}
\end{equation}
where it is noted that $r < \rout$. The function $\gsm$ can now be \emph{any} analytic function in $(r-\rout)$.

Similarly, at $r=\rout$ the function $\lambda(r)$ must vanish, and its first $n$ derivatives must also vanish there.  (Strictly speaking the value of $n$ in (\ref{eq:gsm}) need not be the same as the value of $n$ in (\ref{eq:gsl}). The patching will be $C^{n}$ where $n$ is the lower of the two values.) Assuming that $\lambda(r)$ is analytic we write it, in the domain  $(\rout-\varepsilon) < r \leq \rout$, as
\begin{equation}
 \lambda(r)=\left(\rout-r\right)^{n+1}\gsl \,, \label{eq:gsl}
\end{equation}
where again $\gsl$ can now be \emph{any} analytic function. Therefore in the outer layers of the black hole mimicker we have that in metric (\ref{eq:kerrdsmet})
\begin{equation}
 M\to M-\left(\rout-r\right)^{n+1}\gsm M\, \quad \mbox{and} \quad \Lambda \to \left(\rout-r\right)^{n+1}\gsl \Lambda\,. \label{eq:outerlayerMandL}
\end{equation}

Writing the metric functions as (\ref{eq:gsm}) and (\ref{eq:gsl}) gives us the advantage that they are only subject to the mild restriction of analyticity, and any further properties can be set by the requirement that energy conditions be respected. Further, under this assumption we have that the Einstein tensor and, subsequently via (\ref{eq:einsteqn}) the stress-energy tensor, is analytic, allowing us to expand it in the neighborhood of $\rout$. To lowest non-zero order the orthonormal stress-energy tensor components can be found in Appendix II. 
Specifically we note the following which is of interest: The stress energy components $T_{\hat{1}\hat{1}}$ and $T_{\hat{2}\hat{3}}$ appear at higher order in $(\rout-r)$ than the remaining components. Therefore, in the extreme outer layer of the atmosphere these components can be neglected in comparison to the others.

We consider now the weak energy condition (WEC), the dominant energy condition (DEC), and the strong energy condition (SEC) in the outer atmosphere. As a reminder, these energy conditions read:
\begin{subequations}
\begin{align}
& \forall\;  \boldsymbol{v} \;|\;  v^{\hat{\alpha}} v^{\hat{\beta}} \eta_{\hat{\alpha}\hat{\beta}} < 0\,, \quad T_{\hat{\alpha}\hat{\beta}}v^{\hat{\alpha}}v^{\hat{\beta}} \geq 0\,,  & \mbox{(WEC)} \label{eq:wecgen} \\
& \forall\;  \boldsymbol{v} \;|\;  v^{\hat{\alpha}} v^{\hat{\beta}} \eta_{\hat{\alpha}\hat{\beta}} \leq 0\,, \quad T_{\hat{\alpha}\hat{\beta}}v^{\hat{\alpha}}v^{\hat{\beta}} \geq 0\,,  & \mbox{(DEC)} \label{eq:decgen} \\
& \forall\;  \boldsymbol{v} \;|\;  v^{\hat{\alpha}} v^{\hat{\beta}} \eta_{\hat{\alpha}\hat{\beta}} \leq 0\,, \quad T_{\hat{\alpha}\hat{\beta}}v^{\hat{\alpha}}v^{\hat{\beta}} \geq \frac{1}{2}T^{\hat{\gamma}}_{\;\;\hat{\gamma}} v^{\hat{\delta}} v_{\hat{\delta}} \,.  & \mbox{(SEC)} \label{eq:secgen}
\end{align}
\end{subequations}

In principle one can compute the expressions in (\ref{eq:wecgen})-(\ref{eq:secgen}) using arbitrary timelike and null vectors, whichever is required by the particular energy condition, and try to find if the above weak inequalities are respected for {all} timelike and/or null vectors (which can be parameterized in terms trigonometric and hyperbolic functions). This would be rather complex and instead we note the following: If the stress-energy tensor can be locally diagonalized, then the analysis of the above energy conditions can be sharpened considerably as follows:
\begin{subequations}
\begin{align}
& {_{\tinyrmsub{(d)}}{T}}_{\hat{0}\hat{0}} \geq 0 \;\; \mbox{and}\;\;  {_{\tinyrmsub{(d)}}{T}}_{\hat{0}\hat{0}}+ {_{\tinyrmsub{(d)}}{T}}_{\hat{i}\hat{i}}\geq 0 \quad \mbox{(no sum over $i$)} \,,  & \mbox{(WEC)} \label{eq:wecdiag} \\
& {_{\tinyrmsub{(d)}}{T}}_{\hat{0}\hat{0}} \geq |_{\tinyrmsub{(d)}}{T}_{\hat{i}\hat{i}}| \quad \mbox{(no sum over $i$)}\,,  & \mbox{(DEC)} \label{eq:decdiag} \\
& {_{\tinyrmsub{(d)}}{T}}_{\hat{0}\hat{0}}+ {_{\tinyrmsub{(d)}}{T}}_{\hat{i}\hat{i}}\geq 0 \quad \mbox{(no sum over $i$)} \quad \mbox{and}\quad  {_{\tinyrmsub{(d)}}{T}}_{\hat{0}\hat{0}} + \sum_{\hat{i}} {_{\tinyrmsub{(d)}}{T}}_{\hat{i}\hat{i}} \geq 0\,.  & \mbox{(SEC)} \label{eq:secdiag}
\end{align}
\end{subequations}
where the subscript ``(d)'' means ``diagonalized'', which is related to the Segre characteristic of the tensor \cite{ref:dasdebbook}.

From the discussions above, we note that in the atmosphere the structure of the stress-energy tensor is
\begin{equation}
\left[T_{\hat{\alpha}\hat{\beta}}\right]= \left[\begin{array}{cccc}
                                      \undertilde{T}_{\hat{0}\hat{0}} & 0 & 0 & \undertilde{T}_{\hat{0}\hat{3}} \\
                                      0 & 0 & 0 & 0 \\
                                      0 & 0 & \undertilde{T}_{\hat{2}\hat{2}} & 0 \\
                                     \undertilde{T}_{\hat{3}\hat{0}} & 0 & 0 & \undertilde{T}_{\hat{3}\hat{3}}
                                     \end{array}\right] + \mathcal{O} (r-\rout)^{n} \,, \label{eq:atmsetexp}
\end{equation}
where the under-tilde denotes the lowest order contributions to the stress-energy tensor in the atmosphere, which are of order $(r-\rout)^{n-1}$ and $n$ is the continuity level of the patching between the shell and the Kerr metric at $r=\rout$. Specifically, the quantities $\undertilde{T}_{\hat{\alpha}\hat{\beta}}$ are tabulated in the Appendix II.

We note from (\ref{eq:atmsetexp}) that at lowest order it is only the azimuthal energy flux which spoils the diagonal nature of the stress-energy tensor. Since we are in an orthonormal frame, we can attempt to perform a boost of the \emph{full} (non-expanded) stress-energy tensor in the direction of the energy flux (i.e. a boost in the $\phi$ direction) in order to eliminate the energy flux by going into its locally co-moving frame. It turns out that this is easier to analyze in terms of velocity as opposed to rapidity and hence we utilize the following Lorentz transformation in order to perform this task:
\begin{equation}
\left[\Lambda^{\hat{\alpha}}_{\;\;\hat{\sigma}'}\right]= \left[\begin{array}{cccc}
                                      \gamma(x) & 0 & 0 & \beta(x)\,\gamma \\
                                      0 & 1 & 0 & 0 \\
                                      0 & 0 & 1 & 0 \\
                                     \beta(x)\, \gamma(x) & 0 & 0 & \gamma(x)
                                     \end{array}\right]  \,, \label{eq:boostmatrix}
\end{equation}
where $\beta(x)$ is the local velocity of the boost and $\gamma(x)$ the usual Lorentz factor, $\gamma(x):=1/\sqrt{1-\beta^{2}(x)}$\,. Specifically we wish to calculate the boosted stress tensor,
\begin{equation}
{_{\tinyrmsub{(d)}}{T}}_{\hat{\sigma}'\hat{\rho}'}= \Lambda^{\hat{\alpha}}_{\;\;\hat{\sigma}'}\,T_{\hat{\alpha}\hat{\beta}}\, \Lambda^{\hat{\beta}}_{\;\;\hat{\rho}'}\,, \label{eq:setboost}
\end{equation}
such that
\begin{equation}
 {_{\tinyrmsub{(d)}}{\undertilde{T}}}_{\hat{0}'\hat{3}'}=0\,. \label{eq:noflux}
\end{equation}
That is, the resulting boosted stress-energy tensor, when expanded about $r=\rout$ should have no energy flux to lowest order, nor should any shears appear at this order.
It turns out that the local boost matrix, (\ref{eq:boostmatrix}), is surprisingly simple. In order to diagonalize the stress-energy tensor it is simply required that  the velocity of the boost take on this form:
\begin{equation}
 \beta(x)=\frac{a \sqrt{a^2 + \rout(\rout-2M)}}{a^2+\rout^2}\sin\theta =:\beta_{0}\sin\theta \,, \label{eq:betasol}
\end{equation}
which sets the restriction
\begin{equation}
 -1 < \frac{a \sqrt{a^2 + \rout(\rout-2M)}}{a^2+\rout^2} < 1\,. \label{eq:betarestriction}
\end{equation}
The quantity (\ref{eq:betasol}) is gotten via using (\ref{eq:boostmatrix}) in (\ref{eq:setboost}) and solving for $\beta(x)$ by demanding that (\ref{eq:noflux}) holds.
Since $\rout > r_{\tinyrmsub{H}}$, with $r_{\tinyrmsub{H}}$ given by (\ref{eq:kerrhor}), the quantity under the square root  in (\ref{eq:betasol}) is always positive for $M > |a|$. Also, since the boost has the structure (\ref{eq:boostmatrix}) and also does not depend on $r$ it turns out that, after boosting, the boost does not introduce the radial pressure or a shear at lowest-order in the new frame.

The locally boosted stress-energy components, to lowest order in $(\rout-r)$, are given by (we henceforth drop the primes on the indices since the $(\mathrm{d})$ subscript denotes that we are in the locally diagonal frame):
\begin{subequations}
\begin{align}
{_{\tinyrmsub{(d)}}{\undertilde{T}}}_{\hat{0}\hat{0}}&=\frac{n(n+1)}{12}\, \frac{a^{2}\Lambda\gslrout \left( a^{2}+\rout (\rout-2M) \right)}{2\pi \left(\rout^{2}+a^{2}\cos^{2}\theta\right)} \left(\rout-r\right)^{n-1}\,, \label{eq:rhoatmosp} \\
{_{\tinyrmsub{(d)}}{\undertilde{T}}}_{\hat{2}\hat{2}}&=\frac{n(n+1)}{48\pi \left(\rout^{2}+a^{2}\cos^{2}\theta\right)} \left[6\gsmrout M \rout -\gslrout\Lambda \left(4a^{2}+ \rout^{4}+a^2\rout(5\rout-8M)\right)\right. \nonumber \\
& \left.\quad + a^2\gslrout \Lambda  \left(a^2 +\rout(\rout-2M)\right)\cos^{2}\theta\right] \left(\rout-r\right)^{n-1} \,, \label{eq:pangatmosph} \\
{_{\tinyrmsub{(d)}}{\undertilde{T}}}_{\hat{3}\hat{3}}&=\frac{n(n+1)}{48\pi \left(\rout^{2}+a^{2}\cos^{2}\theta\right)}\left[6\gsmrout M\rout -\gslrout\Lambda \left(2a^{4}+\rout^{4}+a^{2}\rout(3\rout-4M)\right) \right. \nonumber \\
&\left. \quad -a^2\gslrout \Lambda \left(a^{2}+\rout(\rout-2M)\right)\cos^{2}\theta\right] \left(\rout-r\right)^{n-1}\,, \label{eq:pang2atmosph}
\end{align}
\end{subequations}
with all others zero.

\subsection{The weak energy condition (WEC)}
Let us consider the weak energy condition (\ref{eq:wecdiag}) utilizing the stress-energy components (\ref{eq:rhoatmosp}) - (\ref{eq:pang2atmosph}). The study of the energy conditions are greatly facilitated by the following substitutions:
\begin{equation}
 \rout \to r_{\tinyrmsub{H}}+\epsilon\,, \quad \mbox{and} \quad r \to \rout-\delta\,, \label{eq:econdreparams}
\end{equation}
with $\epsilon$ and $\delta$ small and positive, enforcing the conditions that the outer surface of the transition shell is outside of the Kerr horizon, and that the domain in question (the atmosphere) is inside of $\rout$. The Kerr horizon is given by the expression (\ref{eq:kerrhor}) and with parameterizations (\ref{eq:econdreparams}) the square root in (\ref{eq:betarestriction}) reads
\begin{equation}
 \sqrt{a^2 + \rout(\rout-2M)} = \left[\epsilon\left(2\sqrt{M^2-a^2}+\epsilon\right)\right]^{1/2}\,. \label{eq:betasqrt}
\end{equation}

With the substitutions (\ref{eq:econdreparams}), (\ref{eq:rhoatmosp}) becomes
\begin{equation}
 {_{\tinyrmsub{(d)}}{\undertilde{T}}}_{\hat{0}\hat{0}}= \frac{n(n+1)}{24 \pi} \frac{a^2 \delta^{n-1}\epsilon \gslrout \Lambda (\epsilon +2 \sqrt{M^2 - a^2})}{(\epsilon+M+\sqrt{M^2-a^2})^{2} + a^{2}\cos^{2}\theta}\,, \label{eq:weccond1}
\end{equation}
which can be made non-negative by choosing $\gslrout \geq 0$. (Recall that the only restriction on $\gsl$ is that it is an analytic function.) Therefore, the first part of the WEC (\ref{eq:wecdiag}), ${_{\tinyrmsub{(d)}}{\undertilde{T}}}_{\hat{0}\hat{0}} \geq 0$,  can be respected, and this also automatically satisfies ${_{\tinyrmsub{(d)}}{\undertilde{T}}}_{\hat{0}\hat{0}}+ {_{\tinyrmsub{(d)}}{\undertilde{T}}}_{\hat{1}\hat{1}}\geq 0$ since ${_{\tinyrmsub{(d)}}{\undertilde{T}}}_{\hat{1}\hat{1}}$ vanishes at lowest-order. We remind the reader that, via (\ref{eq:econdreparams}), the $r$ dependence now is in $\delta=\rout-r$, which gives the coordinate distance within the transition layer from its outer surface.

We next turn our attention to the next part of the energy condition, ${_{\tinyrmsub{(d)}}{\undertilde{T}}}_{\hat{0}\hat{0}}+ {_{\tinyrmsub{(d)}}{\undertilde{T}}}_{\hat{2}\hat{2}}$. Using (\ref{eq:econdreparams}) in (\ref{eq:rhoatmosp}) and (\ref{eq:pangatmosph})
\begin{align}
 {_{\tinyrmsub{(d)}}{\undertilde{T}}}_{\hat{0}\hat{0}}+ {_{\tinyrmsub{(d)}}{\undertilde{T}}}_{\hat{2}\hat{2}}&= \frac{n(n+1)\delta^{n-1}}{48\pi \left[\left(\epsilon +M +\sqrt{M^{2}-a^{2}}\right)^{2} +a^{2}\cos^{2}\theta\right]}  \times \nonumber \\
 &\Biggl\{ 6\gsmrout M \left(\epsilon +M +\sqrt{M^2-a^2}\right) - \gslrout \Lambda \biggl( 2a^{4}+ \left( \epsilon + M +\sqrt{M^{2}-a^{2}}\right)^{4}  \nonumber \\
&+  a^{2}\left(\epsilon + M + \sqrt{M^{2}-a^{2}}\right)\left(3\epsilon - M + 3\sqrt{M^2-a^2}\right)\biggr) \nonumber \\
& + a^{2}\epsilon \gslrout \Lambda \left(\epsilon +2 \sqrt{M^2-a^2}\right)\cos^{2}\theta \Biggr\}\,. \label{eq:weccond2}
\end{align}
The above will be non-negative provided the term in braces is non-negative. This demands that
\begin{align}
 \gsmrout & \geq \frac{\gslrout \Lambda}{6M\left(\sqrt{M^2-a^2}+\epsilon+M\right)} \nonumber \\
 &\Biggl\{4 \epsilon^3
   \left(\sqrt{M^2-a^2}+M\right)+12 \epsilon^2 M
   \left(\sqrt{M^2-a^2}+M\right)-a^2 \left(-2
   \epsilon \left(\sqrt{M^2-a^2}-5 M\right) \right. \nonumber \\
  & \left. +2 M
   \left(\sqrt{M^2-a^2}+3 M\right)+3
   \epsilon^2\right)-a^2 \epsilon \cos^2\theta  \left(2
   \sqrt{M^2-a^2}+\epsilon\right) \nonumber \\
   &  +16 \epsilon M^2
   \left(\sqrt{M^2-a^2}+M\right)+ 8 M^3
   \left(\sqrt{M^2-a^2}+M\right)+\epsilon^4\Biggr\}\,. \label{wecineq1}
\end{align}

Finally we write out the final part of the WEC
{\begin{align}
 {_{\tinyrmsub{(d)}}{\undertilde{T}}}_{\hat{0}\hat{0}}+ {_{\tinyrmsub{(d)}}{\undertilde{T}}}_{\hat{3}\hat{3}}&= \frac{n(n+1)\delta^{n-1}}{48\pi \left[\left(\epsilon +M +\sqrt{M^{2}-a^{2}}\right)^{2} +a^{2}\cos^{2}\theta\right]}  \times \nonumber \\
 &\Biggl\{ -\left(\sqrt{M^2-a^2}+\epsilon+M\right) \biggl[\gslrout
   \Lambda  \left(\sqrt{M^2-a^2}+\epsilon+M\right)
   \left(\left(\sqrt{M^2-a^2}+\epsilon+M\right)^2+a ^2\right) \nonumber \\
& -6 \gsmrout M\biggr]-a^2 \epsilon \gslrout
   \Lambda  \cos^2\theta \left(2\sqrt{M^2-a^2}+\epsilon\right)  \Biggr\}\,,\label{eq:weccond3}
\end{align}}
which in order to be non-negative demands
\begin{align}
 \gsmrout & \geq \frac{\gslrout \Lambda}{6M\left(\sqrt{M^2-a^2}+\epsilon+M\right)} \nonumber \\
 & \Biggl\{ 4 \epsilon^3
   \left(\sqrt{M^2-a^2}+M\right)+12 \epsilon^2 M
   \left(\sqrt{M^2-a^2}+M\right)-a^2 \biggl[2
   \epsilon \left(\sqrt{M^2-a^2}+5 M\right)  \nonumber \\
   &+2 M
   \left(\sqrt{M^2-a^2}+3 M\right)+5
   \epsilon^2\biggr]+a^2 \epsilon \cos^2\theta \left(2
   \sqrt{M^2-a^2}+\epsilon\right)\nonumber \\
   & +16 \epsilon M^2
   \left(\sqrt{M^2-a^2}+M\right)+8 M^3
   \left(\sqrt{M^2-a^2}+M\right)+\epsilon^4  \Biggr\}\,.
\end{align}
Therefore, large values of $\gsmrout$ will tend to satisfy the WEC. We will reserve commenting further on the results of the weak energy condition until studying the dominant energy condition, since if the DEC can be respected then the WEC is also respected.

\subsection{The dominant energy condition (DEC)} \label{sec:dec}
In order to study (\ref{eq:decdiag}) we shall consider ratios of the pressures, ${_{\tinyrmsub{(d)}}{\undertilde{T}}}_{\hat{i}\hat{i}}$, to the energy density, ${_{\tinyrmsub{(d)}}{\undertilde{T}}}_{\hat{0}\hat{0}}$, noting that we have already established that the energy density is non-negative if $\gslrout \geq 0$. Therefore, throughout the analysis we shall assume
$\gslrout \geq 0$. One should apply caution when simply taking the ratio of two leading independent expansion terms. However, what we wish to illustrate specifically is that  (\ref{eq:decdiag}) holds, and this ratio will allow us to determine that.  If the magnitudes of these ratios are all less than 1 then the DEC is satisfied. The ratios are as follows from (\ref{eq:rhoatmosp})-(\ref{eq:pang2atmosph}):
\begin{subequations}
\allowdisplaybreaks{\begin{align}
\frac{{_{\tinyrmsub{(d)}}{\undertilde{T}}}_{\hat{2}\hat{2}}}{{_{\tinyrmsub{(d)}}{\undertilde{T}}}_{\hat{0}\hat{0}}}&= \frac{1}{2(a^2\epsilon \gslrout\Lambda(\epsilon +2 \sqrt{M^{2}+a^{2}})} \Biggr\{ 6 \gsmrout M \left(\epsilon + M +\sqrt{M^{2}-a^{2}}\right) \nonumber \\
& -\gslrout\Lambda\left(4a^{4} + (\epsilon + M +\sqrt{M^{2}-a^{2}})^{4} \right. \nonumber \\
& \left. + a^{2}(\epsilon + M +\sqrt{M^{2}-a^{2}})(5 \epsilon -3M +5 \sqrt{M^2-a^2})\right)\Biggr\} +\frac{1}{2} \cos^{2}\theta\,, \label{eq:dec1} \\
\frac{{_{\tinyrmsub{(d)}}{\undertilde{T}}}_{\hat{3}\hat{3}}}{{_{\tinyrmsub{(d)}}{\undertilde{T}}}_{\hat{0}\hat{0}}}&= \frac{1}{2(a^2\epsilon \gslrout\Lambda(\epsilon +2 \sqrt{M^{2}+a^{2}})} \Biggr\{ 6 \gsmrout M \left(\epsilon + M +\sqrt{M^{2}-a^{2}}\right) \nonumber \\
& -\gslrout\Lambda\left(2a^{4} + (\epsilon + M +\sqrt{M^{2}-a^{2}})^{4} \right. \nonumber \\
& \left. + a^{2}(\epsilon + M +\sqrt{M^{2}-a^{2}})(3 \epsilon -M +3 \sqrt{M^2-a^2})\right)\Biggl\} -\frac{1}{2} \cos^{2}\theta\,. \label{eq:dec2}
\end{align}}
\end{subequations}
The $1/2\,\cos^{2}\theta$ terms in the above expressions are, of course, bounded by $0$ (at the equator) and $1/2$ (at the poles), and the remaining parts of the expressions do not contain $\theta$. Therefore in the above expressions we must have the following conditions in order to respect the DEC:
\begin{subequations}
\begin{align}
\frac{1}{2} &\geq \frac{1}{2(a^2\epsilon \gslrout\Lambda(\epsilon +2 \sqrt{M^{2}+a^{2}})} \Biggr\{ 6 \gsmrout M \left(\epsilon + M +\sqrt{M^{2}-a^{2}}\right) \nonumber \\
& -\gslrout\Lambda\left(4a^{4} + (\epsilon + M +\sqrt{M^{2}-a^{2}})^{4} \right. \nonumber \\
& \left. + a^{2}(\epsilon + M +\sqrt{M^{2}-a^{2}})(5 \epsilon -3M +5 \sqrt{M^2-a^2})\right)\Biggr\}\geq -1 \,, \label{eq:decb1} \\
1 &\geq  \frac{1}{2(a^2\epsilon \gslrout\Lambda(\epsilon +2 \sqrt{M^{2}+a^{2}})} \Biggr\{ 6 \gsmrout M \left(\epsilon + M +\sqrt{M^{2}-a^{2}}\right) \nonumber \\
& -\gslrout\Lambda\left(2a^{4} + (\epsilon + M +\sqrt{M^{2}-a^{2}})^{4} \right. \nonumber \\
& \left. + a^{2}(\epsilon + M +\sqrt{M^{2}-a^{2}})(3 \epsilon -M +3 \sqrt{M^2-a^2})\right)\Biggl\} \geq -\frac{1}{2}\,. \label{eq:decb2}
\end{align}
\end{subequations}
We notice that, unsurprisingly, the ratios (\ref{eq:dec1}) and (\ref{eq:dec2}) are independent of the radius $r$. (i.e. they are independent of $\delta$) and of the continuity level $n$. We also notice in these expressions that there exists a term that depends on the ratio $\gsmrout/\gslrout$, and a term that depends on neither $\gsmrout$ or $\gslrout$. Therefore, what is important in satisfying the DEC is this ratio of the two constants, and we can write
\begin{equation}
\frac{\gsmrout}{\gslrout}=\alpha_{0}, \quad \mbox{or}\quad \gsmrout=\alpha_{0}\gslrout\,, \nonumber
\end{equation}
where $\alpha_{0}$ is some constant, whose restriction only comes from satisfying the inequalities (\ref{eq:dec1}) and (\ref{eq:dec2}) which should be between $\pm 1$ in order to satisfy the DEC. We remind the reader here that we have already established that $\gslrout > 0$ from the demand that ${_{\tinyrmsub{(d)}}{\undertilde{T}}}_{\hat{0}\hat{0}}$ be non-negative. Solving for $\alpha_{0}$, subject to the restriction that (\ref{eq:dec1}) and (\ref{eq:dec2}) are bounded by $\pm 1$, is rather trivial provided the other quantities are set. We provide an example below in section \ref{sec:atmosph}. By satisfying these DEC restrictions, along with positivity of the energy density, which has already been established, the WEC will also be satisfied.

\subsection{Strong energy condition (SEC)}
Finally, we comment here on the strong energy condition (\ref{eq:secdiag}). Specifically, let us analyze the second expression in (\ref{eq:secdiag}):
\allowdisplaybreaks{\begin{align}
{_{\tinyrmsub{(d)}}{\undertilde{T}}}_{\hat{0}\hat{0}} + \sum_{\hat{i}} {_{\tinyrmsub{(d)}}{\undertilde{T}}}_{\hat{i}\hat{i}} & = \frac{n(n+1)\,\delta^{n-1}}{24 \pi
   \left[\left(\sqrt{M^2-a^2}+\epsilon+M\right)^2+a^2 \cos^2\theta \right]} \Biggl\{-\gslrout \Lambda \Biggl[2 a^4   \nonumber \\
 &  +a^2 \left(\sqrt{M^2-a^2}+\epsilon+M\right) \left(3\sqrt{M^2-a^2}+3\epsilon-M\right) \nonumber \\
  &+\left(\sqrt{M^2-a^2}+\epsilon+M\right)^4\Biggr]+6 \gsmrout M \left(\sqrt{M^2-a^2}+\epsilon+M\right)\Biggr\} \label{eq:sencond}
\end{align}}
Note in the above that a sufficiently large $\gsmrout$ will render the above non-negative. Also, from (\ref{eq:pangatmosph}) and (\ref{eq:pang2atmosph}), sufficiently large $\gsmrout$ also yields non-negative ${_{\tinyrmsub{(d)}}{\undertilde{T}}}_{\hat{2}\hat{2}}$ and ${_{\tinyrmsub{(d)}}{\undertilde{T}}}_{\hat{3}\hat{3}}$, since the contribution to the pressures from the $\gsmrout$ term is positive. Finally, we note also from the earlier discussion of the WEC that $\gslrout \geq 0$ (but not so large as to violate energy conditions) renders ${_{\tinyrmsub{(d)}}{\undertilde{T}}}_{\hat{0}\hat{0}} \geq 0$, so the SEC can in principle be satisfied. Again we illustrate this below in section \ref{sec:atmosph}.

\subsection{A specific atmosphere satisfying all conditions} \label{sec:atmosph}
The analyses above are rather involved, subject to many conditions. It is therefore instructive to show that all of the conditions that have been derived can indeed be satisfied. We consider here the atmospheric region of the black hole mimicker, so that the expansions above hold (subject of course to non-zero radius of convergence). For the illustrative example here we chose the following sets of parameters, although any set satisfying the above conditions will suffice:
\begin{align}
n&=3,           &  M&=1,               &  a&=\frac{7}{8},   &   \Lambda&=\frac{1}{2}, \label{eq:atmosparams} \\
\gslrout&=1\times 10^{6},  & \beta_{0}&=0.129,         &  \alpha_{0}&=0.594,    &  \rout&=r_{\tinyrmsub{H}}+\Scale[0.95]{\frac{1}{2}[M-\sqrt{M^2-a^2}]} \approx 1.742\,. \nonumber
\end{align}
The value of $\beta_{0}$ above arises from using (\ref{eq:betasol}) in (\ref{eq:boostmatrix}) and (\ref{eq:setboost}) and demanding (\ref{eq:noflux}). The value of $\alpha_{0}$ is chosen as one of (infinitely) many values that satisfy the DEC, which admittedly is a fine-tuning issue however, since not all values of $\alpha_{0}$ will allow the DEC to be respected.

With the above parameters, the now non-existent pure Kerr black hole horizon would have been located at $r_{\tinyrmsub{H}}=1.484$. Also, the now non-existent de~Sitter horizon would occur at $r_{\tinyrmsub{dS}}=2.288$ at the equator and at $r_{\tinyrmsub{dS}}=2.449$ at the poles. On the equator, the existing outer ergosurface is located at $r=2$ and this ergosurface crosses into the shell at $\theta=0.698\,$ and $\,\theta=2.443$. Therefore this specific case corresponds to a model where part of the ergoregion is outside of the transition shell and part of it is inside, very similar to what is illustrated in figure \ref{fig:rot_grav_2}. Of course, since we are interested in the atmosphere region, we will consider the domain where $\delta\;(=\rout-r)$ is small and non-negative, and specifically in what follows we will concentrate on the outer $5\%$ of coordinate radius of the gravastar.

\begin{figure}[htbp]
\begin{center}
\includegraphics[width=1.00\textwidth, clip, keepaspectratio=true]{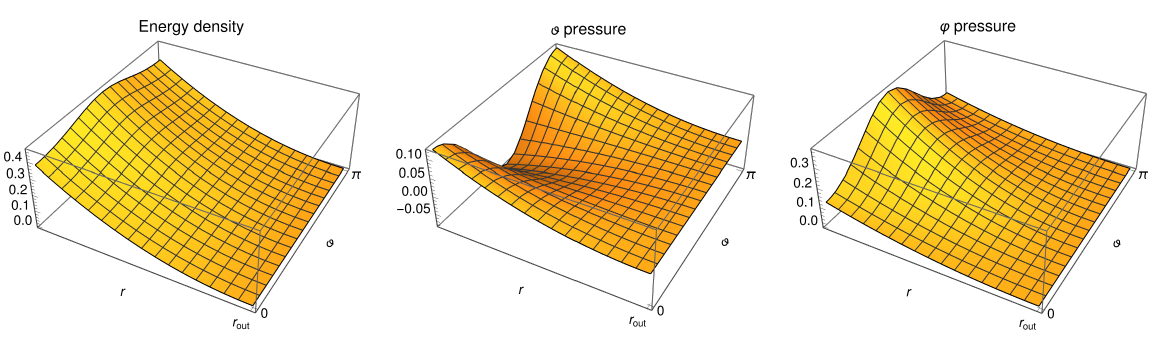}
\caption{{\small{The energy density (left), altitudinal pressure (middle), and azimuthal pressure (right) in the atmosphere of the black hole mimicker.}}}
\label{fig:atmosset}
\end{center}
\end{figure}

We first illustrate the energy density and surviving principle pressures in the outer atmosphere. This is shown in figure \ref{fig:atmosset} where the graphs represent the energy density, altitudinal pressure, and azimuthal pressure respectively. Notice that near the equator the altitudinal pressure is negative.

\begin{figure}[htbp]
\begin{center}
\includegraphics[width=1.00\textwidth, clip, keepaspectratio=true]{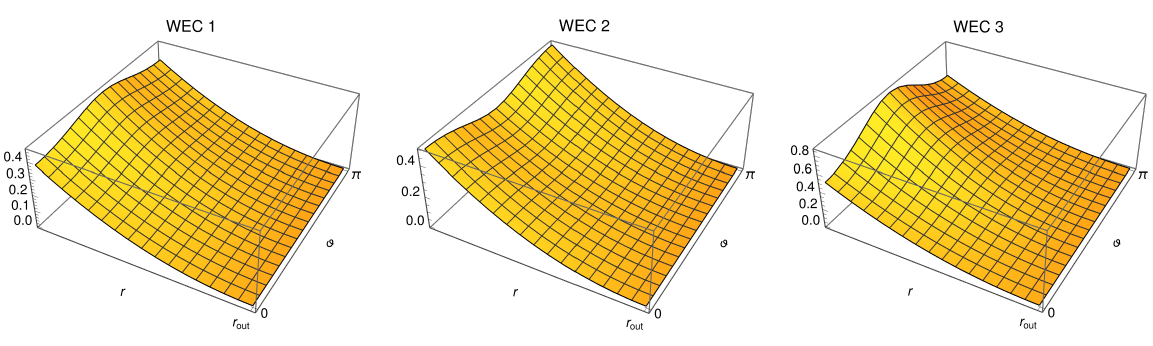}
\caption{{\small{The weak energy condition (WEC) furnished by (\ref{eq:weccond1}), (\ref{eq:weccond2}) and (\ref{eq:weccond3}), respectively, in the atmosphere. All quantities are non-negative}}}
\label{fig:atmoswec}
\end{center}
\end{figure}

In figure \ref{fig:atmoswec} we plot (\ref{eq:weccond1}), (\ref{eq:weccond2}) and (\ref{eq:weccond3}) respectively. It can be noticed that all of these are non-negative illustrating that the WEC is satisfied in the atmosphere.

\begin{figure}[htbp]
\begin{center}
\includegraphics[width=0.80\textwidth, clip, keepaspectratio=true]{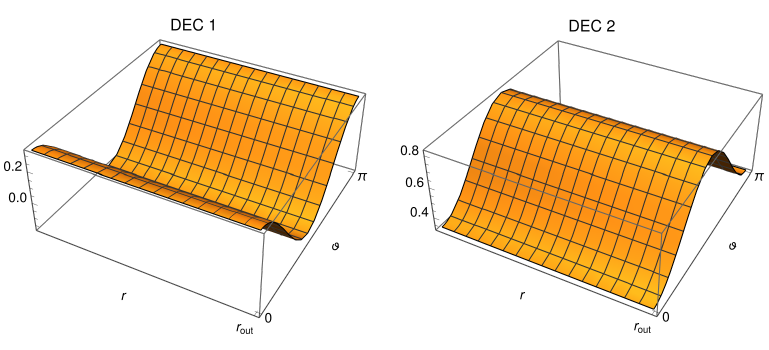}
\caption{{\small{Part of the dominant energy condition (DEC) furnished by (\ref{eq:dec1}) and (\ref{eq:dec2}) respectively, in the atmosphere. These quantities are both bounded by $\pm 1$.}}}
\label{fig:atmosdec}
\end{center}
\end{figure}

In figure \ref{fig:atmosdec} we illustrate the DEC conditions (\ref{eq:dec1}) and \ref{eq:dec2}) respectively. It is noted that both of these are bounded by $\pm 1$ as required for respecting the DEC.

\begin{figure}[htbp]
\begin{center}
\includegraphics[width=0.40\textwidth, keepaspectratio=true]{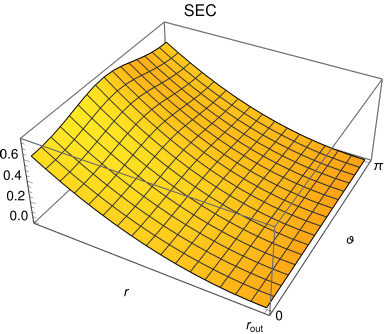}
\caption{{\small{Part of the strong energy condition (SEC) given by (\ref{eq:sencond}) in the atmosphere. This quantity is everywhere non-negative.}}}
\label{fig:atmossec}
\end{center}
\end{figure}

The final energy condition we study is the SEC. We show the condition (\ref{eq:sencond}) within the atmosphere in figure \ref{fig:atmossec}. We can see that this is everywhere non-negative.

In summary, it can be seen that the choice (\ref{eq:atmosparams}) provides one set of parameters which satisfy the weak, dominant, and strong energy conditions in the atmosphere region.

\begin{figure}[H]
\begin{center}
\includegraphics[width=1.00\textwidth, clip, keepaspectratio=true]{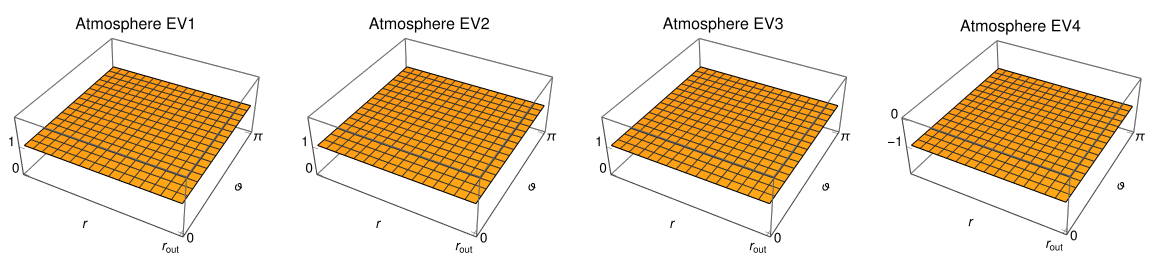}
\caption{{\small{The four normalized eigenvalues of the metric in the atmosphere region. Note that there are three positive and one negative eigenvalue, denoting Lorentzian structure.}}}
\label{fig:atmosevals}
\end{center}
\end{figure}

Finally, we need to show that we possess Lorentzian structure in this atmosphere, which is perhaps not obvious due to the metric components changing various signs (see figure \ref{fig:rot_grav_2}). Admittedly, since we are in general relativity, and energy conditions are respected, it would be somewhat surprising if the spacetime possesses peculiar signature. However, it should be checked for certainty. We do this in figure \ref{fig:atmosevals} where we plot the four eigenvalues (\ref{eq:eval1}) - (\ref{eq:eval4}) divided by their respective magnitudes, using the parameters (\ref{eq:atmosparams}) in the outer region of the gravastar. These show that the signature is $+2$ everywhere in the outer atmosphere.

It is of interest to see what sort of matter the above stress-energy tensor may correspond to. We note from fig.~\ref{fig:atmosset} that when the magnitude of the energy density is large, the magnitude of the two pressures is large, and similarly when the magnitude of the energy density is small, the magnitudes of the two pressures is small. This leads us to believe that a reasonably simple relationship may exist between the energy density and the pressures. The following observations can be made regarding the material's properties in the atmosphere: Generally, it can be shown that for parameters satisfying the energy conditions, we have $p_{\theta}+p_{\phi}= \omega\,\rho$ with $\omega$ equal to $2/3$ or slightly less and this is constant for all values of $r$ and $\theta$. For the particular parameters chosen in the illustrative model here, we have the relationship that $p_{\theta}+p_{\phi}=0.651\rho$.  In general, at the poles the transverse pressures of energy condition respecting material become equal, $p_{\theta}=p_{\phi}$, and also at the poles it can be shown that the equation of state relating each individual pressure to the energy density satisfies $p_{i}=\omega\rho$ with the constant $\omega$ equal to $1/3$ or slightly less. That is, at the poles, the material is highly relativistic. For the specific set of parameters used to generate the plots here we have $p_{i}=0.325\rho$ at the poles. The fact that the $\phi$ pressure near the equator is outwardly increasing in the atmosphere might be a concern from a stability perspective, as mentioned in section \ref{sec:rotgrava},  and a more in-depth analysis would be required, since in the case with rotation the angular momentum of the star can help support the star against instability. (This is also why the radial pressure in the atmosphere at this order can vanish, but the star remains stationary.)

\subsection{Observational distinction from Kerr black hole} \label{sec:obs}
Here we briefly discuss the qualitative observational differences between the gravastar presented above and the pure Kerr black hole. One way to discern this difference is by the analysis of null geodesics in the vicinity of the atmosphere. Outside of the gravastar the geometry is pure Kerr in this model. However, if light rays from lensed background stars enter the atmosphere which, for example like in our sun, is somewhat transparent, we should  be able to observe differences between a true Kerr black hole and a gravastar with an atmosphere.  The atmosphere metric presented above is rather complicated, and the evolution of null geodesics becomes rather computationally intense. An in-depth study of this is planned for a future work. However, to complete this study of the rotating gravastar we will comment here on the lensing features of a simple class of null geodesics which reside in the equatorial plane. 

\begin{figure}[t!]
\begin{center}
\includegraphics[width=\textwidth, clip, keepaspectratio=true]{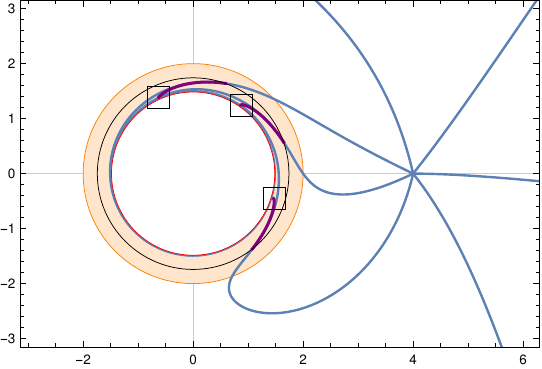}\\[-0.4cm]
\includegraphics[width=0.23\textwidth, clip, keepaspectratio=true]{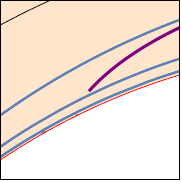} \hfill
\includegraphics[width=0.23\textwidth, clip, keepaspectratio=true]{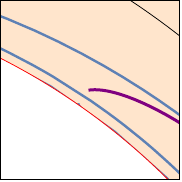} \hfill
\includegraphics[width=0.23\textwidth, clip, keepaspectratio=true]{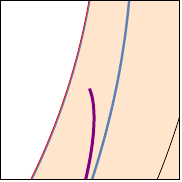}\\[-0.2cm]
\caption{{\small{The evolution of null geodesics from a point light source near the gravastar. Note that in the atmosphere (the region inside the black circle), the gravastar geodesics (purple) are more strongly attracted to the interior than the Kerr geodesics (blue). Insets: A close up of the bifurcation points where the gravastar geodesics deviate from the Kerr geodesics.}}}
\label{fig:nullgeo}
\end{center}
\end{figure}

In figure \ref{fig:nullgeo} we consider a point light source in the vicinity of the gravastar and evolve the null geodesic equation for increasing affine parameter within the equatorial plane. In the figure, the orange shaded region represents the ergoregion of the {pure Kerr black hole}. The red inner-circle represents the event horizon of the pure Kerr black hole, and the black circle represents the outer surface of the gravastar, $\rout$. Note that once the light rays enter the atmosphere they eventually bifurcate into two rays. One ray (blue) represents that light ray as if it were still traveling in the pure Kerr spacetime. The other ray (purple) represents the same geodesic ray except it is for the gravastar atmosphere geometry. We note the following qualitative feature of this bifurcation: The rays propagating in the atmosphere tend to turn inwards more strongly than their corresponding pure Kerr counterparts. Therefore there will be some rays which in the Kerr spacetime will propagate, fully or partially, around the black hole and exit to infinity, whereas the corresponding ray in the gravastar will not make it out to infinity, but instead will plunge into the interior of the gravastar. This feature would in principle allow for an observational distinction between a gravastar and a Kerr black hole. We caution the reader that this plot is in Boyer-Lindquist coordinates where the Boyer-Lindquist radial coordinate is being treated as distance from the center. However, the qualitative feature of stronger attraction is immune to this coordinate effect.

\section{{\small Concluding remarks}}\label{sec:conc}
A scheme has been devised in order to create a class of black hole mimickers, such as a gravastar, with a true de~Sitter interior and a true Kerr exterior. The metric is everywhere $C^{n}$ where $n$ could be made arbitrarily large. The resulting object can be everywhere Lorentzian and non-singular. Depending on how close the location of the exterior surface of the transition region, $\rout$, is to the location of the corresponding Kerr black hole event horizon, the object may mimic a Kerr black hole arbitrarily closely to the event horizon, and it would be difficult to distinguish such an object from a true Kerr black hole via local exterior experiments or observations. The ergoregion's topology is changed in comparison to the truly Kerr geometry, from being bounded by two $S^{2}$ surfaces to being bounded by a single $S^{1}\times S^{1}$ surface. It has also been shown how the outer layers of the black hole mimicker, the atmosphere, can obey the dominant/weak and strong energy conditions. We note that in the equatorial plane null geodesics are more strongly attracted in the atmosphere than their pure Kerr counterparts, leading to an observational distinction between the two objects. Other modes of distinction could involve the study of ringdown effects as well as the study of non-equatorial geodesics. Of particular interest would be light rays propagating closer to the poles where the absence of an ergoregion in the gravastar would likely provide a large difference in behavior compared to the Kerr spacetime.

The non perturbative nature of the construction may open up an avenue for the study of such rotating black hole mimickers that does not depend on various parameters being small (which is sometimes difficult to define in relativity).

\section*{{\small Acknowledgments}} 
This work was partially supported by the VIF program of the University of Zagreb. 

\appendix
\setcounter{equation}{0}
\renewcommand\theequation{A.\arabic{equation}}

\section*{Appendix I- Eigenvalues of the transition region metric}
We present here the eigenvalues of the metric. We first review here a few issues with eigenvalues of the metric tensor and covariance. 

The equation (\ref{eq:evaleqn}) is \emph{not} covariant, and the eigenvalues given by that equation are not true invariants. However, the \emph{number} of positive, negative and zero eigenvalues is invariant. We will refer to these eigenvalues as the ``usual eigenvalues'', as they are the ones most familiar from linear algebra. As well, the units of the metric components, being generally different, means that in the expressions for the usual eigenvalues we will be sometimes adding quantities with different units. In some sense then, the usual eigenvalues themselves are not a useful quantity, but the number of positive, negative, and zero usual eigenvalues is useful, as it yields the signature of the spacetime, and from Sylvester's law this will not change via a change of basis. There do exist also ``Lorentz invariant eigenvalues'' (see, for example \cite{ref:dasdebbook}), although they are of limited use when applied to the metric, though are useful when applied to other quantities which are projected into an orthonormal frame, but we do not need them here as the usual eigenvalues fulfill our purpose. (See also \cite{ref:dim4}.) \\[0.1cm]

The transition shell metric possesses the following usual eigenvalues:
\begin{subequations}
\begingroup
\allowdisplaybreaks
\begin{align}
\ell_{(1)}&= 3\frac{a^{2}\cos^{2}\theta + r^{2}}{\lambda(r)\Lambda a^2\cos^{2}\theta +3 } \,, \label{eq:Aeval1} \\
\ell_{(2)}&= -3\frac{a^{2}\cos^{2}\theta + r^{2}}{\lambda(r)\Lambda r^2 (a^2+r^2)+6\mu(r)M r -3 ( a^2 +r^2)} \,, \label{eq:Aeval2} \\
\ell_{(3)/(4)}&=\frac{1}{2(r^2+a^2\cos^{2}\theta)(\lambda(r)\Lambda a^2+3)^{2}} \times \nonumber \\
&\Scale[0.95]{\Biggl\{ 18 \left(-\frac{a^{2} \Lambda \left(a^{2}+r^{2}+1\right) \lambda \left(r \right)}{6}+M \mu \! \left(r \right) r -\frac{a^{2}}{2}-\frac{r^{2}}{2}\right) a^{2} \cos^{4}\theta }  \nonumber \\
&\Scale[0.95]{+\left(3 a^{2} \Lambda \left(a^{4}-r^{4}+a^{2}\right) \lambda \! \left(r \right)
- 36 M \mu \! \left(r \right) a^{2} r +9 a^{4}-9 r^{4}-9 a^{2}\right) \cos^{2}\theta  } \nonumber \\
&\Scale[0.95]{+3 \Lambda \,r^{2} \left(a^{2}+1\right) \left(a^{2}+r^{2}\right) \lambda \! \left(r \right)+18 M r \left(a^{2}+1\right) \mu \! \left(r \right)+9 a^{2} r^{2}+9 r^{4}-9 r^{2} } \nonumber \\
&\Scale[0.95]{ \mp   3 \biggl[ 36 M^{2} r^{2} \left(a^{2} \cos^{4}\theta-2 a^{2} \cos^{2}\theta+a^{2}+1\right)^{2} \mu \! \left(r \right)^{2}-12 \left(r^{2}+a^{2} \cos^{2}\theta\right) M r \left(\left(a^{4} \left(a^{2}+r^{2}-1\right) \cos^{6}\theta  \right. \right. } \nonumber \\
 &\Scale[0.95]{+\left(\left(-3 a^{4}-3 a^{2}\right) r^{2}-3 a^{6}-a^{4}\right) \cos^{4}\theta+\left(\left(3 a^{4}+5 a^{2}\right) r^{2}+3 a^{6}+4 a^{4}+a^{2}\right) \cos^{2}\theta } \nonumber \\
 &\Scale[0.95]{-\left.   \left(a^{2}+1\right)^{2} \left(a^{2}+r^{2}\right)\right) \Lambda \lambda \! \left(r \right)} \nonumber \\
 & \Scale[0.95]{+3 \left.\left(a \cos^{2}\theta-a +1\right) \left(a \cos^{2}\theta-a -1\right) \left(\left(a^{2}+r^{2}\right) \cos^{2}\theta-a^{2}-r^{2}-1\right)\right) \mu \! \left(r \right) } \nonumber \\
&\Scale[0.95]{+\left(r^{2}+a^{2} \cos^{2}\theta\right)^{2} \left(\left(a^{2} \left(\left(a^{2}+4\right) r^{4}+\left(2 a^{4}+6 a^{2}\right) r^{2}+a^{2} \left(a^{2}+1\right)^{2}\right) \cos^{4}\theta  \right.\right. } \nonumber \\
&\Scale[0.95]{ -2 \left. \left(\left(a^{2}+3\right) r^{2}+\left(a^{2}+1\right)^{2}\right) a^{2} \left(a^{2}+r^{2}\right) \cos^{2}\theta+\left(a^{2}+1\right)^{2} \left(a^{2}+r^{2}\right)^{2}\right) \Lambda^{2} \lambda \! \left(r \right)^{2} } \nonumber \\
&\Scale[0.95]{+6 \left(\left(a^{2}+r^{2}-1\right) a^{2} \cos^{2}\theta +\left(-a^{2}+1\right) r^{2}-a^{4}+a^{2}\right) \left(\left(a^{2}+r^{2}\right) \cos^{2}\theta-a^{2}-r^{2}-1\right) \Lambda \lambda \! \left(r \right)} \nonumber \\
&\Scale[0.95]{ +9 \left. \left(\left(a^{2}+r^{2}\right) \cos^{2}\theta-a^{2}-r^{2}-1\right)^{2}\right) \biggr]^{1/2}\Biggr\} }\,. \label{eq:Aeval34}
\end{align}
\endgroup
\end{subequations}


\section*{Appendix II- Leading order contribution to the orthonormal stress-energy tensor in the outer layer}
We present here the lowest-order contribution to the orthonormal stress-energy tensor of the transition region. We only show the lowest-order contribution due to the complication of the full expressions. However, we should emphasize that when we boost the stress-tensor in the main text it is the \emph{full} tensor that is boosted, and the expansion is only performed after boosting, which leads to (\ref{eq:rhoatmosp}) - (\ref{eq:pang2atmosph}). (This is the mathematically cleaner procedure; boosting followed by expansion.) 
\begin{subequations}
{\allowdisplaybreaks\begin{align}
&\undertilde{T}_{\hat{0}\hat{0}}= \nonumber \\
&\frac{n(n+1)}{48 \pi  \left(a^2 \cos (2 \theta )+a^2+2 r_{\text{out}}^2\right){}^2 \
\left(4 a^2 M r_{\text{out}} \sin ^2\theta+\left(a^2+r_{\text{out}}^2\right) \left(a^2 \cos (2 \theta )+a^2+2  r_{\text{out}}^2\right)\right)}  \nonumber \\
&\times a^2 \left(a^2+\rout \left(\rout-2 M\right)\right) \left(a^2 \cos ^2\theta+\rout^2\right) \biggl[\Lambda  \gslrout \biggl(7 a^4+a^2 \cos (4 \theta ) \left(a^2+\rout \left(\rout-2 M\right)\right)  \nonumber \\
&  +a^2 \rout \left(18 M+19 \rout\right)+4 \cos (2 \theta ) \left(2 a^4+a^2 \rout \left(3 \rout-4 \
M\right)+\rout^4\right)+12 \rout^4 \biggr) \nonumber \\
& +48 \gsmrout M \rout \sin ^2\theta\biggr] (\rout-r)^{n-1}\;, \\[0.2cm]
&\undertilde{T}_{\hat{2}\hat{2}}= \frac{n(n+1)}{48 \pi \left(a^2 \cos (2 \theta )+a^2+2 \rout^2\right)} \biggl[\Lambda  \gslrout \biggl(-7 a^4+a^2 \cos (2 \theta ) \
\left(a^2+\rout \left(\rout-2 M\right)\right) \nonumber \\
& +a^2  \rout \left(14 M-9 \rout\right)-2 \rout^4\biggr)+12 \gsmrout M \rout\biggr] (\rout-r)^{n-1}\;, \\[0.2cm] 
&\undertilde{T}_{\hat{3}\hat{3}}= \frac{n(n+1)}{24 \pi  \left(a^2 \cos (2 \theta )+a^2+2 \rout^2\right) \left(a^4+a^2 \cos (2 \theta ) \left(a^2+\rout \left(\rout-2 M\right)\right)+a^2 \rout \left(2 M+3 \rout\right)+2 \rout^4\right)} \nonumber \\
&\times \biggl[\Lambda  \gslrout \biggl(-3 a^8+a^6 \rout \left(2 M-13 \rout\right)+a^4 \rout^2 \left(8 M^2+12 M \rout-19 \rout^2\right)+a^2 \rout^5 \left(10 M-11 
\rout\right) \nonumber \\
&-a^2 \cos (2 \theta ) \left(a^2+\rout \left(\rout-2 M\right)\right) \left(3 a^4+4 a^2 \rout \left(\rout-M\right)+\rout^4\right)-2  \rout^8\biggr) \nonumber \\
&+12 \gsmrout M \rout \left(a^2+\rout^2\right){}^2\biggr] (\rout-r)^{n-1}\;, \\[0.2cm] 
&\undertilde{T}_{\hat{0}\hat{3}}=\frac{n(n+1)}{24 \pi  \left(a^2 \cos (2 \theta )+a^2+2 \rout^2\right) \left(a^4+a^2 \cos (2 \theta ) \left(a^2+\rout \left(\rout-2 M\right)\right)+a^2 \rout \left(2 M+3 \rout\right)+2 \rout^4\right)} \nonumber \\
& \times \biggl[ a \left(a^2+\rout^2\right) \sqrt{\sin ^2\theta \left(a^2+\rout \left(\rout-2 M\right)\right)} 
\biggl(\Lambda  \gslrout \biggl(a^4+a^2 \cos (2 \theta )  \left(a^2+\rout \left(\rout-2 M\right)\right) \nonumber \\
& +a^2 \rout \left(3 \rout-2 M\right)+2 \rout^4\biggr)-12 \gsmrout M \rout\biggr)\biggr](\rout-r)^{n-1}\;.
\end{align}}
\end{subequations}

\PRLsep

\clearpage


\linespread{0.6}
\bibliographystyle{unsrt}

\end{document}